\documentstyle[11pt, aaspp4]{article}

\begin{document}
\newcommand{\cc}{\mbox{cm$^{-3}$}}
\def\kms{km s$^{-1}$}
\newcommand{\tauv}{\mbox{$\tau_V$}}
\newcommand{\ra}{\mbox{$\rightarrow$}}
\newcommand{\nhtwo}{\mbox{n$_{H_{2}}$}}

\def\HI{H{\smc I}}
\def\HII{H{\smc II}}
\def\m17{M~17}               
\def\cepa{Cepheus~A}               
\def\Htwo{H$_2$}               
\def\HtwoO{H$_2$O}             
\def\HtwoCO{H$_2$CO}           
\def\HtwoCS{H$_2$CS}           
\def\Hthreep{H$_3^+$}          
\def\HtwoDp{H$_2$D$^+$}          
\def\HCOp{HCO$^+$}             
\def\DCOp{DCO$^+$}             
\def\HthCOp{H$^{13}$CO$^+$}    
\def\HCeiOp{HC$^{18}$O$^+$}    
\def\HtwCsiOp{H$^{12}$C$^{16}$O$^+$} 
\def\HCSp{HCS$^+$}             
\def\HthCN{H$^{13}$CN}         
\def\HCfiN{HC$^{15}$N}         
\def\HtwCfoN{H$^{12}$C$^{14}$N}  
\def\HNthC{HN$^{13}$C}         
\def\HfoNtwC{H$^{14}$N$^{12}$C}  
\def\HCthreeN{HC$_3$N}         
\def\twCO{$^{12}$CO}           
\def\thCO{$^{13}$CO}           
\def\CseO{C$^{17}$O}           
\def\CeiO{C$^{18}$O}           
\def\twCsiO{$^{12}$C$^{16}$O}  
\def\thCsiO{$^{13}$C$^{16}$O}  
\def\twCeiO{$^{12}$C$^{18}$O}  
\def\thCeiO{$^{13}$C$^{18}$O}  
\def\CtfS{C$^{34}$S}           
\def\thCS{$^{13}$CS}           
\def\twCttS{$^{12}$C$^{32}$S}  
\def\tfSO{$^{34}$SO}           
\def\ttSsiO{$^{32}$S$^{16}$O}  
\def\SOtwo{SO$_2$}             
\def\tfSOtwo{$^{34}$SO$_2$}    
\def\Ntwo{N$_2$}               
\def\Otwo{O$_2$}               
\def\NtwoHp{N$_2$H$^+$}        
\def\NHthree{NH$_{3}$}         
\def\CHthreeCCH{CH$_3$C$_{2}$H}     
\def\CHthreeCN{CH$_3$CN}       
\def\CHthreeOH{CH$_3$OH}       
\def\CHfour{CH$_4$}       
\def\COtwo{CO$_2$}       
\def\thCHthreeOH{$^{13}$CH$_3$OH}       
\def\twCHthsiOH{$^{12}$CH$_3$$^{16}$OH} 
\def\CtwoH{C$_2$H}             
\def\CHp{CH$^{+}$}             
\def\Cp{C$^+$}             
\def\CthreeHtwo{C$_3$H$_2$}    
\def\Jthoh{$J = 3/2 \to 1/2$}
\def\Johoh{$J = 1/2 \to 1/2$}
\def\Jtwel{$J = 12 \to 11$}
\def\Jelt{$J = 11 \to 10$}
\def\Jtn{$J = 10 \to 9$}
\def\Jne{$J = 9 \to 8$}
\def\Jes{$J = 8 \to 7$}
\def\Jss{$J = 7 \to 6$}
\def\Jsf{$J = 6 \to 5$}
\def\Jff{$J = 5 \to 4$}
\def\Jft{$J = 4 \to 3$}
\def\Jtt{$J = 3 \to 2$}
\def\Jto{$J = 2 \to 1$}
\def\Joz{$J = 1 \to 0$}
\def\WCO{W({\rm CO})}
\def\Wtw{W({\rm ^{12}CO})}
\def\Wth{W({\rm ^{13}CO})}
\def\dv{\Delta v}
\def\dvtw{\Delta v({\rm ^{12}CO})}
\def\dvth{\Delta v({\rm ^{13}CO})}
\def\NCO{N({\rm CO})}
\def\Nth{N({\rm ^{13}CO})}
\def\Ntw{N({\rm ^{12}CO})}
\def\NtwCsiO{N({\rm ^{12}C^{16}O})}
\def\NthCO{N({\rm ^{13}CO})}
\def\NthCsiO{N({\rm ^{13}C^{16}O})}
\def\NtwCeiO{N({\rm ^{12}C^{18}O})}
\def\intCO{\int T_R({\rm CO})dv}
\def\inttwCsiO{\int T_R({\rm ^{12}C^{16}O})dv}
\def\intthCsiO{\int T_R({\rm ^{13}C^{16}O})dv}
\def\inttwCeiO{\int T_R({\rm ^{12}C^{18}O})dv}
\def\NHtwo{N({\rm H_2})}
\def\Wtw{W_{12}}
\def\Wth{W_{13}}
\def\kappanu{\kappa_{\nu}}
\def\phinu{\varphi_{\nu}}
\def\taunu{\tau_{\nu}}
\def\dv{\Delta v}
\def\dvFWHM{\Delta v_{FWHM}}
\def\vLSR{v_{LSR}}
\def\Rsol{R_\odot}
\def\Msol{M_\odot}
\def\MMsol{\ts 10^6\ts M_\odot}
\def\MCO{M_{\rm CO}} 
\def\Mvir{M_{\rm vir}}
\def\TAstar{T^*_A}
\def\TAstartwCO{T^*_A(^{12}{\rm CO})}
\def\TAstarthCO{T^*_A(^{13}{\rm CO})}
\def\TAstarCeiO{T^*_A({\rm C}^{18}{\rm O})}
\def\TRstar{T^*_R}
\def\TexCO{T_{ex}({\rm CO})}
\def\Trms{T_{rms}}
\def\d{^\circ}
\def\h{^{\rm h}}
\def\mi{^{\rm m}}
\def\s{^{\rm s}}
\def\mum{\ts \mu{\rm m}}
\def\mm{\ts {\rm mm}}
\def\cm{\ts {\rm cm}}
\def\percm{\ts {\rm cm}^{-1}}
\def\m{\ts {\rm m}}
\def\kms{\ts {\rm km\ts s^{-1}}}
\def\K{\ts {\rm K}}
\def\Kkms{\ts {\rm K\ts km\ts s^{-1}}}
\def\kHz{\ts {\rm kHz}}
\def\MHz{\ts {\rm MHz}}
\def\GHz{\ts {\rm GHz}}
\def\pc{\ts {\rm pc}}
\def\kpc{\ts {\rm kpc}}
\def\Mpc{\ts {\rm Mpc}}
\def\cmsq{\ts {\rm cm^2}}
\def\pcsq{\ts {\rm pc^2}}
\def\dsq{\ts {\rm deg^2}}
\def\debye{\ts10^{-18}\ts {\rm esu}\ts {\rm cm}}

\title{THE IONIZATION FRACTION IN DENSE MOLECULAR GAS II: MASSIVE CORES}

\author{Edwin A. Bergin, Ren\'e Plume, Jonathan P. Williams, and Philip C. Myers}

{Harvard-Smithsonian Center for Astrophysics, 
60 Garden St., Cambridge, MA 02138-1596; 
ebergin, rplume, jpw, pmyers@cfa.harvard.edu}\newline

\begin{abstract}
We present an observational and theoretical study of the ionization fraction in several
massive cores located in regions that are currently forming stellar clusters.
Maps of the emission from the J $= 1 \rightarrow 0$ transitions of \CeiO , \DCOp ,
\NtwoHp , and \HthCOp , as well as the J $= 2 \rightarrow 1$ and J $= 3 \rightarrow 2$
transitions of CS, were obtained for each core.   Core densities are determined via
a large velocity gradient analysis with values typically $\sim 10^5$ \cc .  With the
use of observations to constrain variables in the chemical calculations we derive
electron fractions for our overall sample of 5 cores directly associated with
star formation and 2 apparently starless cores.   The electron abundances are found
to lie within a small range, $-$6.9 $<$ log$_{10}$($x_e$) $< -7.3$, and are consistent with
previous work.   We find no difference in the amount of ionization fraction between cores 
with and without associated star formation activity, nor is any difference found in electron
abundances between the edge and center of the emission region.  Thus our models are
in agreement with the standard picture of cosmic rays as the primary source of ionization 
for molecular ions.
With the addition of previously determined electron abundances for
low mass cores, and even more massive cores associated with O and B clusters, 
we systematically examine 
the ionization fraction as a function of star formation activity.  This analysis
demonstrates that the most massive sources
stand out as having the lowest electron
abundances $(x_e < 10^{-8}$). 

\end{abstract}


\lefthead{Bergin et al.}
\righthead{Ionization Fraction in High Mass Cores}

\pagebreak

\section{Introduction}

In the standard model of star formation the magnetic field plays the major 
role in the support of dense cores against gravitational contraction (e.g. Shu, Adams,
\& Lizano 1987). 
While the mechanism for the magnetic support may be direct support by the field 
and/or by turbulent motions induced by the field, it is certain
that the magnetic field can only directly influence the motions of
charged species inside the cloud.  Thus the majority of gas mass, residing in 
neutral species, is only influenced via collisions with the ionized species, and 
the level of ionization inside the core becomes a critical factor in the dynamics of the 
star formation process.  

The standard model has predominantly focused on the formation of low mass
stars, because these objects sometimes form in an isolated fashion and, therefore, are
easier to study individually.  In this model the diffusion of magnetic flux inside
the core is controlled by the ionization fraction 
through ion-neutral collisions, or ambipolar diffusion, and the initially
magnetically subcritical core evolves to the supercritical state and collapse
ensues (e.g. McKee et al. 1993).  However, the majority of star formation (both high
and low mass) does not occur in an isolated fashion.  Instead most stars generally form
in clusters or aggregates (Lada \& Lada 1991; Zinnecker, McCaughrean, \& Wilking 1993).  
Thus, the standard model of star formation may not
be applicable to most stars.  In particular, the enhanced radiation fields 
in regions of cluster formation may increase ion abundances and, therefore, the 
coupling of the magnetic field to the gas.  Star formation may, therefore, require
a different formation mechanism.   

To examine the coupling of the magnetic field to the
gas in dense cores we have undertaken an observational and theoretical
study to survey the ionization fraction in dense cores
with a variety of star forming properties.
In the companion paper to this work, Williams et al. (1998; hereafter Paper I), we combined observations of
molecular neutrals and ions with chemical theory and determined 
electron abundances inside an isolated sample of low mass cores.  We found that
the ionization fraction ($x_e = n_e/n_{H_2})$ is typically $\sim 10^{-7}$ and
shows no difference between cores with embedded stars and those without
(see also Caselli et al. 1998).  This value is in agreement 
with previous estimates of
the ionization fraction in Gu\'elin, Langer, \& Wilson (1982) and Wootten et al. (1982), which were based on observations of fewer cores and on less detailed
chemical models. 
In this paper we extend upon the previous work by
deriving the ionization fraction in
cores located in the L1641 and L1630 molecular clouds -- two regions that
are currently forming stellar clusters
(e.g. Lada et al. 1991a; Strom, Strom, \& Merrill 1993).  The molecular cores in
these regions have properties that are typically associated with clouds forming
high mass stars and stellar clusters in that they are generally larger, more massive
and turbulent, than the low mass cores observed in Paper I.
 
Using a set of consistent assumptions and the same chemical model,
we will compare the fractional ionization of these objects with their 
low mass counterparts.   
With the addition of previously determined ionization fractions
in even more massive star forming sites in de Boisanger et al. (1996), and the low
mass results in Paper I we examine
the abundance of electrons throughout most of the range of star formation conditions 
currently found in the galaxy.
In \S 2 we discuss the observations.  In \S 3 we present the integrated intensity maps
for the observed transitions in each source.  Section 4 discusses the physical and chemical
models adopted for these sources and the electron abundances derived from the analysis.
Finally, \S 5 discusses the implications of these results on field-gas coupling 
and on the stability of molecular cores.

\section{Observations}

\subsection{Sources}

The L1641 and L1630 molecular clouds are part of 
Orion A and Orion B respectively.  
Both Orion A and Orion B have been extensively surveyed in CO (Maddalena et al. 1986), 
$^{13}$CO (Bally, Langer, \& Liu 1991), and CS (Lada et al. 1991b; Tatematsu et al. 1993a) 
with each cloud having $\sim 10^5$ M$_{\odot}$ of H$_2$; L1641 alone contains $\sim 2 \times
10^4$ M$_{\odot}$ (Fukui et al. 1986).  These clouds are at a distance of 400 -- 500 pc
and are active sites of star formation.  In L1641
Strom et al. (1993) and Allen (1996) surveyed the stellar population
and suggest that there is a distributed population of young stars
scattered throughout the entire cloud, and also several stellar groups of stars or
aggregates of $\sim 10 - 50$ stars.  In direct contrast to the Trapezium Cluster
in the northern portion of L1641, the population of young stars in the southern
end appears
to be predominantly low to intermediate mass.  A census of young stars in the 
L1630 cloud finds that 
between 54--96\% of the stars in the cloud are grouped into dense stellar clusters
(Lada et al. 1991a).  A more recent J, H, and K photometric search for a 
population of stars on the edges of L1630 finds little evidence for a distributed
population (Li, Evans, \& Lada 1997). 

Within L1641 and L1630
we have chosen sources based upon the IRAS-selected NH$_3$ survey of Harju, Walmsley,
\& Wouterloot (1993; hereafter HWW) who mapped the \NHthree\ $(J,K) = (1,1)$ and
$(2,2)$ emission in dense cores associated with IRAS point sources.
In some instances, while mapping \NHthree\ emission around IRAS
sources, HWW found local maxima in the \NHthree\ emission maps that 
were
apparently unassociated with star formation activity.

We have mapped the emission of 3 molecular ions (\DCOp , \HthCOp , \NtwoHp ) and 
two neutral species (\CeiO , CS) around 
five IRAS sources, containing a total of 8 ammonia cores as defined by
local maximum of the \NHthree\ 
emission mapped by HWW.  For the purposes of this paper we define a ``core'' as 
the position corresponding to a local maximum in the \DCOp\ and \HthCOp\ data and 
not in \NHthree\ emission.
Thus, mostly as a result of our lower angular resolution, 
the targeted 8 HWW ammonia peaks reduce to a total of 7 cores
(5 with stars and 2 without) 
in our sample of molecular ions.   Of these, six are directly associated with the target
HWW sources and one is a newly discovered starless core near IRAS05389--0756 that is 
discussed in \S 3.4.

The central map coordinates are given in 
Table 1.  We have adopted the nomenclature of HWW in our discussion
of source names.  Of the sources contained in our survey all cores 
designated with (a) are associated with IRAS sources and cores with (b,c,...)
are starless cores without any apparent star formation activity. 
The designation of sources as ``starless'' is determined by comparison to the IRAS
point source catalog.   In addition, one source (IRAS05369--0728a,b)
was studied in the J, H, and K-band
survey of Strom et al. (1993) and we have used their observations to perform a deeper
search for associated embedded stars.  We find that all infrared sources coincident
with core (b) have colors consistent with those of 
background stars seen through the edges of the core, although there is a 3.6 cm
continuum source (without a near-IR counterpart) that is on the edge of the 
emission region (Anglada et al. 1992).
The region surrounding core (a) has been discussed in detail by Tatematsu et al. 
(1993b), who argue that there are potentially two young stellar objects associated
with the core.
Except for IRAS05369--0728a,
none of the cores in our survey is directly associated 
with stellar clusters or associations.
However, these sources all have near neighbors that are currently forming clusters or
associations.  

\subsection{Sample Properties and Differences}
 
HWW presented a comparison of the properties of cores found in Orion with those 
in the Taurus molecular cloud.  They demonstrate that the Orion cores have masses,
sizes, and velocity line widths that are more than a factor of two higher than
the Taurus cores.  Among our sample of the HWW Orion cores the average
radius and temperature is 0.16 pc and 15 K, respectively.  The low mass
cores in Paper I have an average radius of $\sim$ 0.09 pc, with an average temperature of
10 K (Benson \& Myers 1989).  
Furthermore, the character of stars forming within our sample of Orion
cores is distinctly different when compared to those in Paper I.  The median infrared 
luminosity of the stars in the Orion sample (as derived from IRAS)
is $\sim 25$ $L_{\odot}$,
as compared to a median of $\sim$3 $L_{\odot}$ for stars associated with the cores observed
in Paper I.  The low mass cores are also primarily associated with one or no star,
while the Orion cores are in a region where cores form single stars or small
groups of $< 30$. 
These characteristics also distinguish our sample of Orion cores 
(and the low
mass cores) from even more massive objects such as OMC-1 in Orion
or W3.  These more massive cores have greater levels of turbulence with typical
line widths of $2-5$ km s$^{-1}$, as compared to $\sim 1$ km s$^{-1}$ for our 
surveyed cores, and $<$ 0.4 km s$^{-1}$ for Taurus cores (Goldsmith 1987; HWW). 
Such regions are also warmer with gas temperatures typically $> 30$ K, and are 
associated with stellar clusters containing $\sim 100 - 1000$ stars,
including luminous O and B stars.  

For the purposes of this paper, following the notation of Myers, Ladd, \& Fuller
(1991) and Caselli \& Myers (1995), we refer to our sample of Orion cores as
``massive'' cores, while those in Paper I are labeled as low mass cores.
This label does not readily distinguish the ``massive'' Orion cores from
the more massive regions, such as OMC-1 in Orion, or W3.  As such when
we refer to OMC-1, W3, or sources with similar characteristics, 
we add the additional qualifier
of ``very'' massive.  
 
\subsection{Molecular Lines}

Beam sampled \HthCOp\ and \DCOp\ J $= 1 \rightarrow 0$
maps of each core 
were obtained at the National Radio Astronomy
Observatory (NRAO) 12m antenna on January 18 -- 21, 1997.    The line center frequencies
are listed in Table 2. In good weather, the dual channel 3mm receiver
achieved system temperatures of 250--300 K.  
The data were obtained using a combination of backends, but we present only the
data from the hybrid correlator operating with 12.5 MHz bandwith and 24.4 kHz channel 
spacing providing velocity resolution of 0.10 km s$^{-1}$ at 72 GHz and 0.08 km s$^{-1}$
at 86 GHz.  To improve the signal to noise these data have been boxcar smoothed
by a factor of two in velocity.
The maps were created on a point by point basis with 75$''$ spacing
in order to sample the emission.  Integration times varied with source and position 
but ranged from 2 to 5 min.  Also at the NRAO on October 8, 1997 we obtained 
single pointed observations of the J = 3 $\rightarrow$ 2 transition of CS towards
the central position of each source.  These data were taken with a system
temperature of $200-250$ K, an integration time of eight minutes per point, and the 100 kHz filterbanks 
with a velocity resolution of 0.05 km s$^{-1}$.  All observations were obtained
using frequency switching with a 6 MHz offset.

During March 1997 each source was mapped in the J $= 1 \rightarrow 0$
transitions of \CeiO\ and \NtwoHp , and the J $= 2 \rightarrow 1$ transition of
CS using the Five College Radio
Astronomy Observatory 14m antenna equipped with the 15-element focal plane array.
The maps were full beam sampled and cover the entire region mapped in \HthCOp\ and
\DCOp\ at NRAO.  We used the autocorrelator spectrometer with
20 MHz bandwith and velocity resolutions of 0.11 km s$^{-1}$, 0.13 km s$^{-1}$,
0.12 km s$^{-1}$ for \CeiO , \NtwoHp , and CS respectively. 
All observations were obtained using frequency switching 
with a 4 MHz throw.   
In the contour maps (Figures 1 -- 5) and for the analysis we convert all data to
$T_{R} = T_A^*/\eta_{MB}$ using the efficiencies given in Table 2 (see
Kutner \& Ulich 1982).

\section{Results}

\subsection{IRAS05399--0121a}

\begin{figure}
\figurenum{1}
\plotfiddle{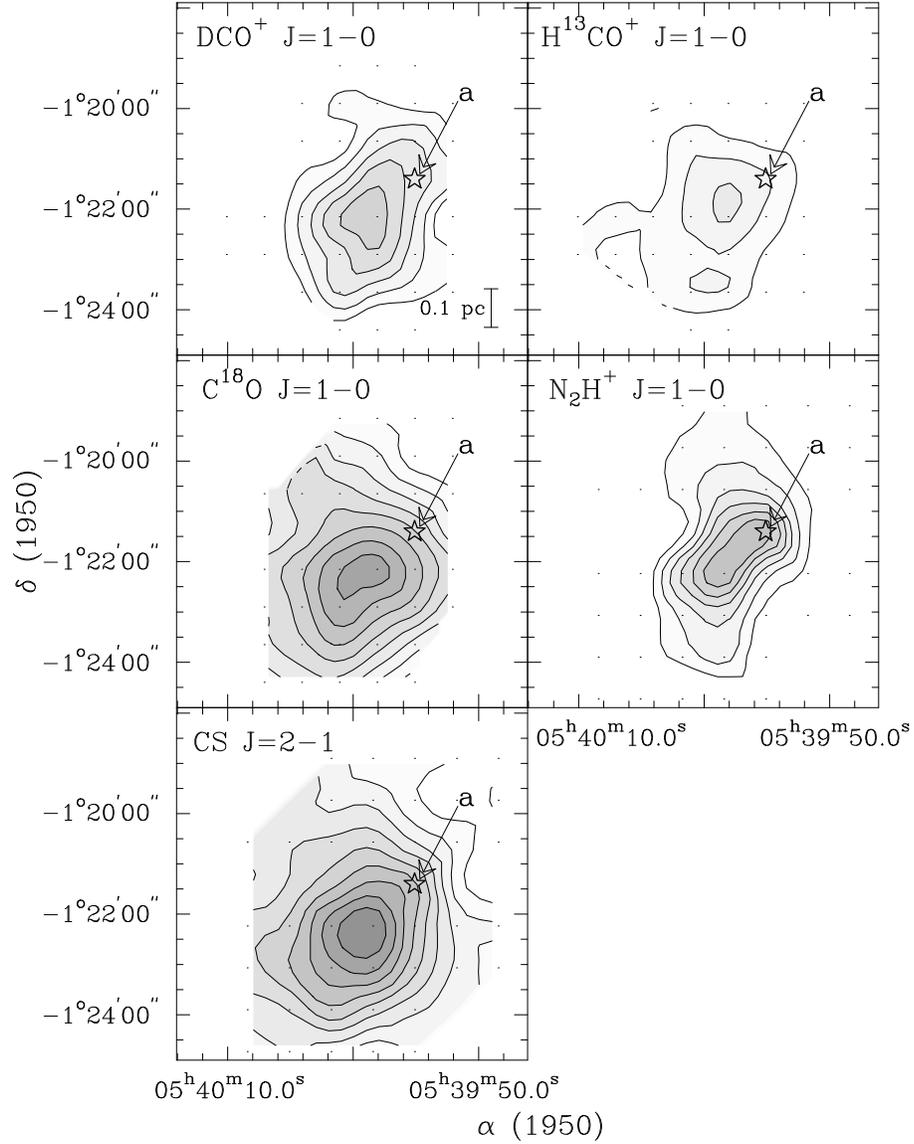}{6in}{0}{70}{70}{-200}{0}
\caption{
Integrated intensity maps ($\int T_R dv$) for the molecular transitions mapped in
IRAS 05399--0121.  The letters denote the position of the NH$_3$ core
as listed by Harju et al. (1993) and the  star symbol represents
the IRAS point source.   In this source the ammonia peak position and
the IRAS source are at the same position. Contour levels and grey 
scale intensities for \DCOp\ and \HthCOp\ 
begin at 0.2  K km s$^{-1}$ and increase in steps
of 0.2  K km s$^{-1}$.   For CS and \NtwoHp\ the contour levels begin at
0.4 K km s$^{-1}$ and increase in steps of  0.4  K km s$^{-1}$.
\CeiO\ contour levels begin at 2.0  K km s$^{-1}$ and increase in steps of 0.4
K km s$^{-1}$.  Peak values are listed in Table~\ref{line_param}.
}
\label{f05399}
\end{figure}

Figure~\ref{f05399} presents integrated intensity maps of \DCOp , \HthCOp , \NtwoHp ,
\CeiO , and CS in the molecular core surrounding IRAS05399--0121.  
The morphology of the emission maps is similar, except that
the \NtwoHp\ emission appears to peak closer to the
star than seen for other molecular species.  
The \CeiO\ integrated intensity map is somewhat larger than observed
in the other tracers, and is most likely due to the lower critical 
density of this transition, n$_{cr} \sim 10^3$ \cc , as compared to
the other tracers which all have n$_{cr} > 10^{5}$ \cc\ (e.g. Ungerechts et al. 1997).
HWW list this core as a possible outflow candidate on the basis
of $^{12}$CO emission widths greater than 10 km s$^{-1}$. Our CS spectra show some slight
evidence for line wings, but there is no other evidence for an outflow in our data.
This core has been observed previously and is listed as
LBS30 in the Lada et al. (1991b) study and 
OriB9 in Caselli \& Myers (1995).   
LBS30 has also been observed in dust continuum emission at both moderate
($\sim 30''$) and high ($\sim 12''$) resolution (Launhardt et al. 1996). 
The moderate resolution maps exhibit a clumpy structure with the strongest emission 
found near the IRAS source and a second weaker peak about $\sim 1'$ to the southeast.
The high resolution data show a single small condensation located within 20$''$ of 
IRAS05399, but within the IRAS error ellipsoid.
The emission extent in our (lower resolution) molecular data encompasses all of the dust
continuum emission.  However, the molecular integrated intensity peaks appear to correspond
with the second weaker dust continuum peak and not with the condensation 
directly associated with the young star.

\subsection{IRAS05302--0537a,b}

\begin{figure}
\figurenum{2}
\plotfiddle{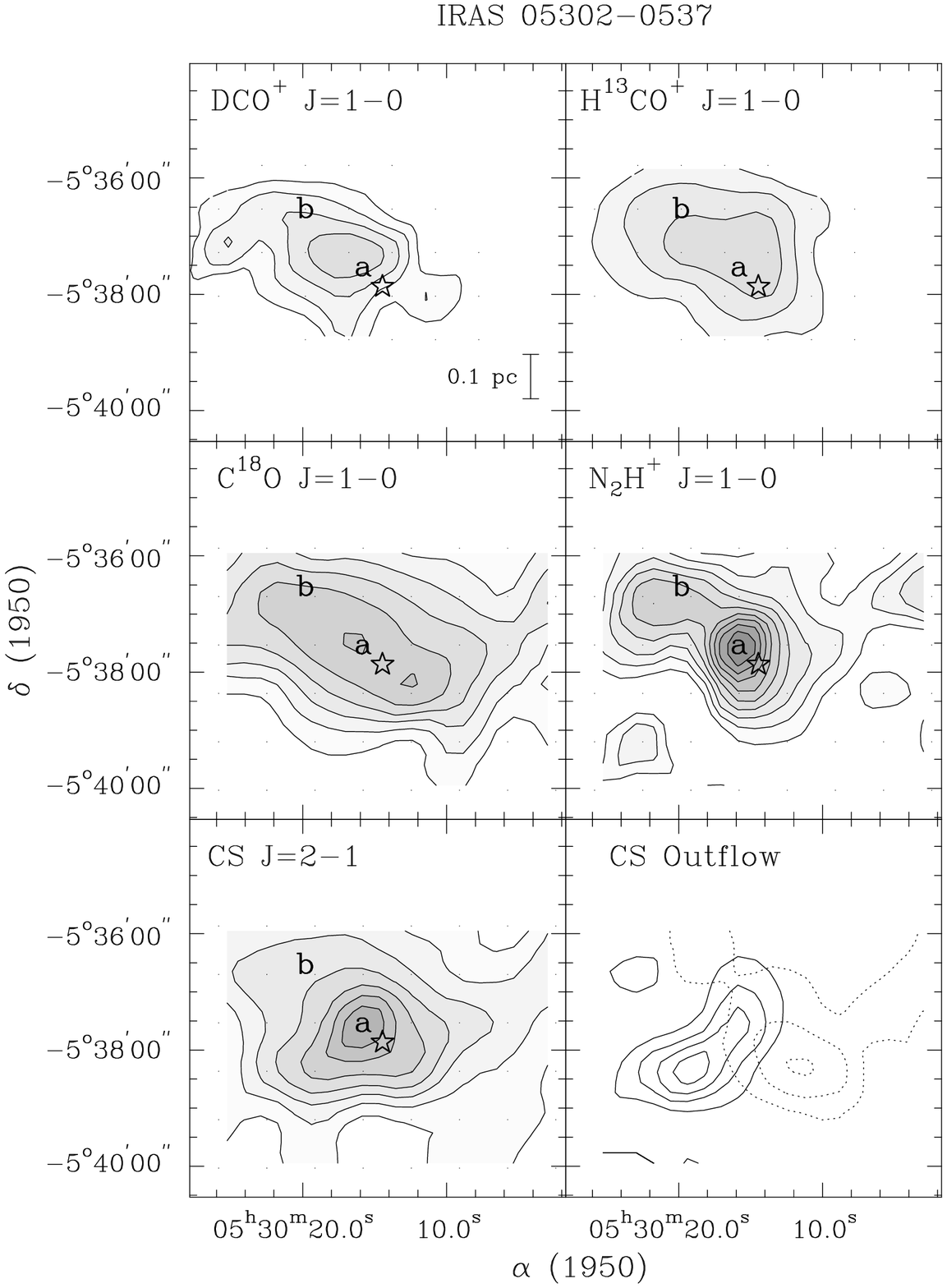}{6in}{0}{70}{70}{-200}{0}
\caption{
Integrated intensity maps ($\int T_R dv$) for various molecular transitions mapped in
IRAS 05302--0537.  Symbols are defined as given in Fig. 1. 
\DCOp\ and \HthCOp\ contour levels and grey 
scale intensities begin at 0.25  K km s$^{-1}$ and increase in steps
of 0.25  K km s$^{-1}$.  For CS and \NtwoHp\ contour levels begin
at 0.5 K km s$^{-1}$ and increase in steps of 0.5  K km s$^{-1}$.
Peak values are given in Table~\ref{line_param}.
Contour levels for both the blue (solid line) and red (dashed line)
outflow lobes begin at 0.2 K km s$^{-1}$ and increase by 0.2 K km s$^{-1}$.
The high velocity blue emission was integrated from 6.4 to 7.6 km s$^{-1}$ and the red
from 9.7 to 11.0 km s$^{-1}$.
}
\label{f05302}
\end{figure}

Contour maps of the integrated intensity from each of the lines observed towards
IRAS05302--0537 are presented in Figure~\ref{f05302}. 
The \NtwoHp\ emission peaks directly to the north of the 
IRAS source and extends to the northeast.  A similar 
distribution is observed for \DCOp\ and \HthCOp .
HWW designate the core to the northeast as a separate 
core (IRAS05302--0537b), but our lower resolution maps do not
resolve a separate source.  However, there are some observed emission differences.
The \CeiO\ distribution encompasses all of the peaks found by other molecules,
while the intensity of the other molecular species are highest near the 
IRAS source. IRAS05302--0537a does have an associated molecular outflow
(Fukui et al. 1986), which is 
observed in the CS emission (shown on the bottom right-hand panel of Figure~\ref{f05302}).  
However, because the outflow contribution to the CS emission is relatively weak it
does not appreciably affect the overall CS distribution.
The young star IRAS05302--0537 is also referred to as Orion A-W and a discussion
of its properties can be found in Meehan et al. (1998).

\subsection{IRAS05369--0728a,b}

\begin{figure}
\figurenum{3}
\plotfiddle{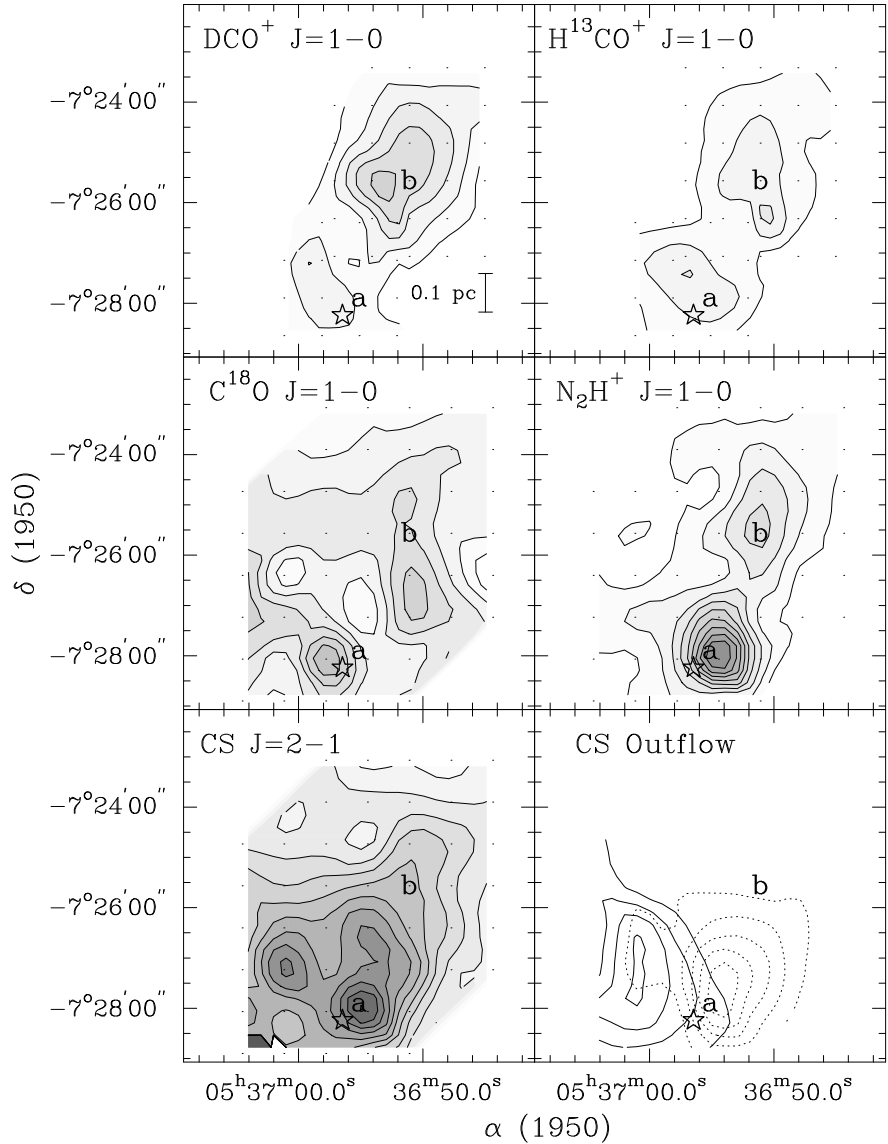}{6in}{0}{70}{70}{-200}{-15}
\caption{
Integrated intensity maps ($\int T_R dv$) for various molecular transitions mapped in
IRAS 05369--0728.  Symbols are defined as given in Fig. 1.
\DCOp\ and \HthCOp\ contour levels and grey 
scale intensities begin at 0.25  K km s$^{-1}$ and increase in steps
of 0.25  K km s$^{-1}$. 
\NtwoHp\ contour levels are 1.0  K km s$^{-1}$ to 10.0 K km s$^{-1}$ by steps
of 1.0  K km s$^{-1}$. CS levels begin at 0.5 K km s$^{-1}$ and increase in
steps of 0.5 K km s$^{-1}$.  Peak values are given in Table~\ref{line_param}.
Contour levels for both the blue (solid line) and red (dashed line)
outflow lobes are 0.4 K km s$^{-1}$ to 4.0 K km s$^{-1}$ by 0.3 K km s$^{-1}$.
The high velocity blue emission was integrated from 2.0 to 3.7 km s$^{-1}$ and the red
from 6.0 to 8.0 km s$^{-1}$.
}
\label{f05369}
\end{figure}

Figure~\ref{f05369} shows maps of the integrated intensity in the molecular gas near
IRAS05369--0728 (also known as Haro 4-255).  
Towards
this source Anglada et al. (1989) and HWW observed two \NHthree\ cores, 
with one sharply peaked condensation 
near the IRAS source and another less defined maximum 2$'$ to the northwest. 
The distribution of our \NtwoHp\ data closely resembles their ammonia observations.
The second maximum away from the IR source is the starless core labeled as IRAS05369-0728b.
For the other molecular ions, the structure also appears to be reasonably 
close to that seen in \NtwoHp . 
In contrast, the observed emission from both neutral species, CS and \CeiO , 
exhibits some differences from the ion distribution.

For CS the differences can be accounted for by the
presence of an outflow from the IRAS source which contributes to 
the CS emission.   An outflow associated with Haro 4-255 
has been reported previously on the basis of CS J $1 \rightarrow 0$
emission (Tatematsu et al. 1993b).  
The red and blue lobes of the outflow as observed by the
J$ = 2 \rightarrow 1$ transition of CS are shown in the bottom right-hand
panel of Figure~\ref{f05369}.  This distribution is quite similar to that found in  
emission from the lower transition  in Tatematsu et al. (1993b).
If we remove the outflow contribution from the
CS integrated intensity map by restricting the integration to the
line core we find that the distribution is comparable to that observed
in the other high dipole moment tracers (\DCOp , \HthCOp , \NtwoHp ).
The morphological differences between \CeiO\ and the other molecules are somewhat
harder to reconcile and are most likely related to the differences in excitation
requirements discussed in \S 3.1.

\subsection{IRAS05389--0756a,b,d}

\begin{figure}
\figurenum{4}
\plotfiddle{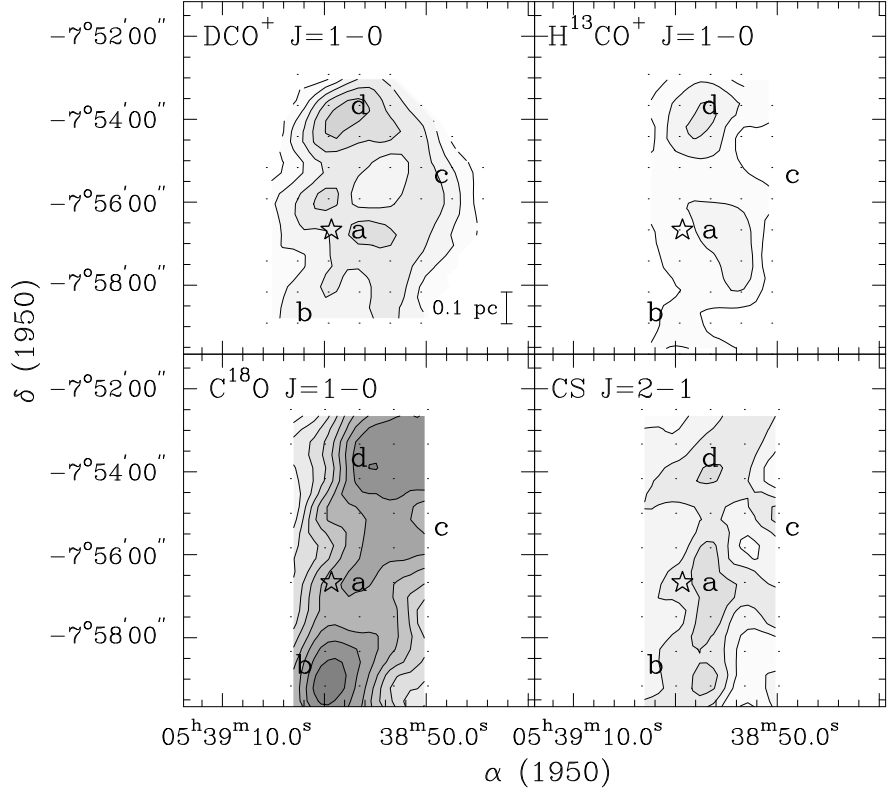}{6in}{0}{70}{70}{-200}{0}
\caption{
Integrated intensity maps ($\int T_R dv$) for various molecular transitions mapped in
IRAS 05389--0756.   Symbols are defined as given in Fig. 1, except for the
open triangle which denotes the DCO$^+$
peak that we call IRAS 05389--0756d in the text. 
Contour levels and grey 
scale intensities for \DCOp\ and \HthCOp\ 
begin at 0.2  K km s$^{-1}$ and increase in steps
of 0.2  K km s$^{-1}$. 
\CeiO\ contour levels begin at 2.0 K km s$^{-1}$ and increase in steps of
0.4 K km s$^{-1}$.  CS and \NtwoHp\ levels begin at 0.4 K km s$^{-1}$ and increase in steps
of 0.4  K km s$^{-1}$.
Peak values are given in Table~\ref{line_param}.
}
\label{f05389}
\end{figure}

Figure~\ref{f05389} presents the intensity maps for surveyed species towards IRAS05389--0756.
For this source we do not have maps of the \NtwoHp\ emission.
While 
HWW detected 3 components in this core, our maps only cover two of these
components (a and b).  In \NHthree\ emission core (a) 
is a spatially broad condensation located near
the star, while core (b) is
a weak \NHthree\ maximum to the south.  
In our observations we see little evidence for the second core,
and it is likely that core (b) is unresolved within our larger beam.  
Comparing the emission morphologies of \HthCOp\ and \DCOp\ we find distributions similar 
to ammonia,
except that \HthCOp\ has an additional maximum 3$'$ north of the IRAS source.
This core was not detected in \NHthree\ emission because 
the maps did not extend this far to the north. 
We label this new starless core as IRAS05389--0756d.

Some of the \CeiO\ and \HthCOp\ spectra in this source show evidence for two 
velocity components centered near 4 and 5 km s$^{-1}$ (see Table 3).  As the spatial
and velocity resolution of the \DCOp\ observations are not high enough to 
resolve these components  we have integrated over the entire line profile
when calculating column densities (\S 4.1.2).

\subsection{IRAS05403--0818a}

\begin{figure}
\figurenum{5}
\plotfiddle{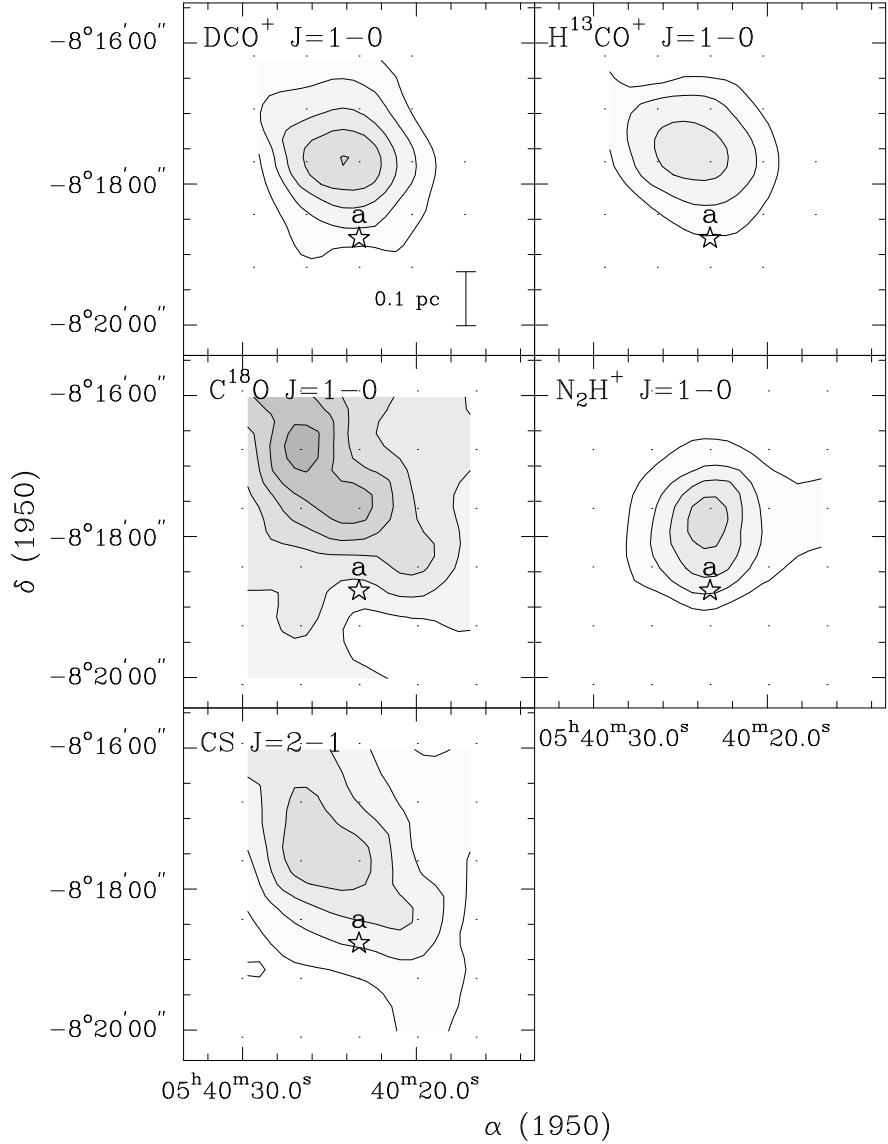}{6in}{0}{70}{70}{-200}{0}
\caption{
Integrated intensity maps ($\int T_R dv$) for various molecular transitions mapped in
IRAS 05403--0818.  Symbols are defined as given in Fig. 1.
For \DCOp\ and \HthCOp\ maps, contour levels and grey 
scale intensities begin at 0.2  K km s$^{-1}$ and increase in steps
of 0.2  K km s$^{-1}$.  All other species begin at 0.4 K km s$^{-1}$ and increase in steps
of 0.4 K km s$^{-1}$.  Peak values are given in Table~\ref{line_param}.
}
\label{f05403}
\end{figure}

Maps of the integrated intensity towards IRAS05403--0818 are presented in Figure~\ref{f05403}.
Again general agreement is observed between the various tracers.  CS, \DCOp , \HthCOp ,
and \NtwoHp\ exhibit a single emission peak $\sim1'$ north of the IRAS source.
The \NHthree\ peak seems to disagree with the peak position observed for
other tracers by $\sim 0.3-0.6'$.  However, closer examination of the ammonia emission distribution
in HWW shows that the peak could also be assigned at $\sim 0.33'$ north, 
close to the peak positions observed in our data.
Similar to the other sources, \CeiO\ is distributed over a larger area and has a more
extended peak.   The \CeiO\ spectra have an additional component at $V_{lsr} \sim 5.7$
km s$^{-1}$, in contrast to the single component at $V_{lsr} \sim 3.0$ km s$^{-1}$ observed in other tracers.
This extra \CeiO\ velocity component is probably due to a background cloud and has not been included in the
intensity map in Figure~\ref{f05403}.   This source is also listed as containing a molecular outflow by
HWW.  Our CS maps show no evidence for any high--velocity gas. 

\section{Analysis}

To derive electron abundances we utilize the 
[\DCOp ]/[\HCOp ] and  [\HCOp ]/[\CeiO ] column density ratios.  
Using these ratios requires reasonably
high signal to noise for all relevant species as the relative errors will add in
quadrature.  In order to obtain the maximum signal to noise,
we restrict our determination of molecular column densities and line parameters
to the \DCOp\ emission peaks for each
core.  
The integrated intensities and line parameters derived toward these 
positions for \CeiO , \DCOp , \HthCOp , and \NtwoHp are given in Table~\ref{line_param},  
and CS in Table~\ref{CSline_param}.
The offsets are relative to the central coordinate provided in Table~\ref{centralpos}. 
The resolution of the \CeiO\ observations were obtained with a 
slightly smaller beam size than those of \HthCOp\ and 
we have smoothed the \CeiO\ observations to a resolution of 65$''$. 
Since we have emission maps for each core -- which contain additional information
on electron abundances at the core edges -- in Section 5.2 we will also examine a
column density computed from averaging the integrated intensities away from the peak.

\subsection{Physical Properties and Molecular Abundances}

As discussed in detail in Paper I, the determination of electron abundances
from observations of molecular ions requires information on the
physical structure of the cloud, in particular the density and temperature
of the gas probed by the respective tracers.  These parameters are
required to constrain the chemical model and also to calculate 
the molecular column densities from the integrated intensities.

\subsubsection{Density and Temperature}

The temperature of these cores, as traced by the symmetric top molecule
\NHthree , is typically $T_k \sim 15$ K (HWW) and we will adopt this 
value for our excitation and initial chemical analysis.  Because the derivation of electron
abundances relies on column density ratios derived from transitions with similar
excitation requirements (see Table~2) our results are not highly sensitive to this
choice.

To determine the density and column density in each core we
have used a Large Velocity Gradient (LVG) code to
calculate the radiative transfer of CS (Snell 1981).  We ran a 20$\times$20
grid of models in density -- column density space and then
fitted our observed CS J = $2\rightarrow1$ and J = $3\rightarrow2$
observations to the grid of models using a $\chi^2$ minimization
routine.  The densities in the grid ranged from $10^4$ to $10^6$
cm$^{-3}$ and the column densities per velocity interval, $N/\Delta V$
from $10^{11}$ to $10^{14}$ cm$^{-2}$ (km s$^{-1}$)$^{-1}$.
We ran two grids: one for T$_k$ = 15 K and one for 25 K to
allow for slightly hotter gas. 
The best fit models are presented in Table~\ref{CSden} and show that 
the densities of the gas probed by CS are typically $\sim 10^{5}$
\cc .    

The density uncertainty is also listed in Table 5 and this value is based
on the observed baseline rms noise, and an assumed systematic uncertainty of 30\%.
One concern is that the density is derived on the basis of two 
adjacent rotational transitions that may not be sensitive to material with
a density much greater than the critical density of the \Jtt\ transition.   
We can attempt to address these questions by comparison to other measurements
in the literature.
One core (IRAS05399) was included in a multitransitional CS study in Lada, Evans,
\& Falgarone (1997) who detected emission from the \Jff\ transition.  They 
derive a density of \nhtwo\ $= 2 \times 10^{5}$ \cc , analogous to our result. 
However, Launhardt et al. (1996) derive a density of $>$ 2.9 $\times 10^{7}$ cm$^{-3}$ on
the basis of 1.3mm dust continuum observations of the same object.  
Another dust continuum study of the gas associated with IRAS05389 derives 
a density for the envelope (r $> 10''$) of $\sim 10^{5}$ \cc\
(Zavagno et al. 1997; S72 in their notation), similar to our value.  
The large difference between CS and dust continuum density measurements in 
IRAS05399 may be
due to the higher resolution of the  dust continuum observations ($\sim 12''$),
which will result in greater sensitivity to warmer, denser,
material close to IRAS05399.  Our lower resolution molecular data ($60 - 80''$) using
transitions with low excitation energies may only
probe outer layers which have lower (and more average) densities.

\subsubsection{Molecular Column Densities and Ratios}

Column density estimates for CS have been made on the basis of the non-LTE excitation
model described in the preceding paragraph.  
For \CeiO , \DCOp , \HthCOp , and \NtwoHp\ we assume
that the emission is optically thin and that the excitation temperature 
is greater than the background.   As in Paper I we initially compute column densities
using LTE.  However, at the densities of these cores, n$_{H_2} \sim 10^5$ \cc , the 
level populations will not be in LTE and the column densities derived via observations
of low--J levels will be overestimated.  We have used a statistical equilibrium model
at n$_{H_2} = 10^5$ \cc\ and $T_k = 15$ K to estimate the non-LTE correction factor 
for each species assuming optically thin emission.   
The non-LTE correction factors are 0.96 for \CeiO , 0.51 for
\DCOp , 0.58 for \HthCOp , and 0.65 for \NtwoHp .
The column densities corrected using the beam efficiencies listed in Table 2, and corrected for non-LTE populations, 
are listed in Table~\ref{colden}.  To estimate the column 
density of \HCOp\ and CO we have used [$^{12}$C]/[$^{13}$C] $= 60$ and [$^{16}$O]/[$^{18}$O]
$= 500$.

In the above analysis we have assumed that the emission is optically
thin.  We have some support for this assumption from observations of
\HCeiOp\ J $= 1 \rightarrow 0$  ($\nu =$ 86.75433 GHz) at 
two
positions in IRAS05369-0728.   These observations were obtained at the \HthCOp\ peak 
and at $\sim 0.75'$ east of that position. 
For these positions we
find 1$\sigma$ detection limits of $\int T_R dv$(\HCeiOp ) $< 0.12$ K km s$^{-1}$ and 
$ < 0.02$ K km s$^{-1}$ respectively. 
Using the isotope ratios listed above,
the opacity of \HthCOp\ is $< 1$ for both positions.   Thus, the \HthCOp\ emission
for this core, and by extension the others, is likely to be thin.
For \CeiO\ and \DCOp\ we have no independent information on the 
opacity.  However, the \DCOp\ and \CeiO\ lines are typically gaussian shaped with 
$T_R < 3.0$ K,
which is below the estimated temperature of 15 K, suggesting that the emission is
thin.   
 
\subsection{Chemical Model}

To derive ionization fractions we have adopted the 
method described in Paper I.  
In particular, we vary both physical parameters and 
atomic abundances  in the chemical model to match observed abundances of
molecular ions and column density ratios and derive electron fractions.  
The chemical model is described in Bergin \& Langer (1997) and
uses the pure gas-phase reaction network from Millar, Farquhar, \& Willacy (1996). 

The physical parameters varied are the density ($n_{H_2}$), gas temperature ($T_{gas}$), 
cosmic ray ionization rate of \Htwo\ ($\zeta_{H_2}$), visual extinction ($A_V$), 
and the enhancement, $\chi$, of the local ultra-violet (UV)
radiation field above the normal interstellar value of 1.6 $\times 10^{-3}$ erg cm$^{-2}$
s$^{-1}$ (Habing 1968). 
The density in these sources is constrained by the 
emission analysis discussed in Section 4.1 to lie within a range of 
$n_{H_2} = 3 - 30 \times 10^4$ \cc , and we adopt $n_{H_2} =
10^5$ \cc\ for our analysis.  
The visual extinction within a \CeiO\ beam can be determined using
the \CeiO\ column densities provided in Table~\ref{colden}.  If we assume that [\CeiO ]/[\Htwo ] $ = 5.0
\times 10^{-7}$ (based on direct H$_2$ and CO measurements in Lacy et al. (1994) and
[$^{16}$O]/[$^{18}$O] = 500), and N(\Htwo ) $= 10^{21}A_V$ then
extinctions range from $A_V = 4 - 12$ mag at the chosen positions.  We have adopted
a value of $A_V = 7.5$ mag for our analysis, which is near the middle of this
range.  On the basis of the ammonia observations we adopt $T_k = 15$ K,
but we allow for the temperature to be as high as
30 K in the chemical calculations.  
Since the cores are located in a region of massive
star formation where the radiation field could be higher than average
we examine changing the UV enhancement factor and the visual extinction
(discussed in \S 5.2).

In Paper I, to constrain the range of potential cosmic ray ionization rates, we 
balanced the heating due to
a variable cosmic ray flux with the molecular line cooling calculations of 
Neufeld, Lepp, \& Melnick (1995) and Goldsmith \& Langer (1977) for a cloud temperature
of 10 K.  A similar analysis allowing for temperatures as high as 15--20 K suggests that
ionization rates between $\zeta_{H_{2}} \sim 1 - 15 \times 10^{-17}$ s$^{-1}$ 
will heat the gas to temperatures that lie within the allowed range. 
We have therefore performed model runs with $\zeta_{H_{2}} =$ 1, 5, 10, and
15 $\times 10^{-17}$ s$^{-1}$.  This was done for all permutations
of varied parameters
(i.e. density, temperature, initial atomic abundances).  From these models we found that 
$\zeta_{H_{2}} = 5 \times 10^{-17}$ s$^{-1}$ best reproduced the observations 
and we have adopted this value for the remainder of this paper.  This
value of $\zeta_{H_{2}}$ is consistent with
the value found in Paper I for low mass cores.

Additional parameters in our model are the initial abundances
of carbon, oxygen, nitrogen, and heavier ``metal'' atoms.   As in Paper I 
we alter the initial chemical abundances from fiducial
values which represent abundances typically used for theoretical chemical models. 
These fiducial conditions have C, O, and N with mild depletions with respect
to solar (factor of 2--3), and
the heaver species (Mg, Fe, S....) depleted by several orders of magnitude from solar
values.  Thus the fiducial set of initial atomic abundances are
already depleted from Solar
and we will raise and lower our initial model abundances from these fiducial values.
We use the same fiducial set of abundances as given in Paper I. 
For each set of parameters (density, temperature, cosmic ray flux,
etc.) the abundances of metal ions are varied from 1/10 to 150 times the normal
value.  For a given cosmic ray ionization rate this will effectively raise and lower the  
ionization level respectively.  
Upon comparison with observations we found that only a small range of factors between 
2 -- 10 times fiducial values is required.  
To account for variable depletions in the carbon and oxygen pools
the carbon and oxygen abundances were also varied from 0.5 to 1.2 
times the fiducial values.  We also varied the nitrogen abundance
and found that it does not affect the derived $x_e$.

In summary, perhaps one of the greatest uncertainties in the calculation of theoretical
chemical abundances is the large number of variables involved in the calculation.
In this work we have attempted to avoid this obstacle by computing over
600 model runs covering as many permutations of the parameter space as possible.
Where possible we have used observations to constrain particular variables.
In the following section we demonstrate that a reasonable range of 
these parameters can cover the observed spread in the data -- and 
allow for an estimation of electron abundances in these cores.  
A compilation of the chosen model parameters for our ``best fit'' model is 
provided in Table~\ref{tab_model}.

\subsection{Electron Abundances} 

\begin{figure}
\figurenum{6}
\plotfiddle{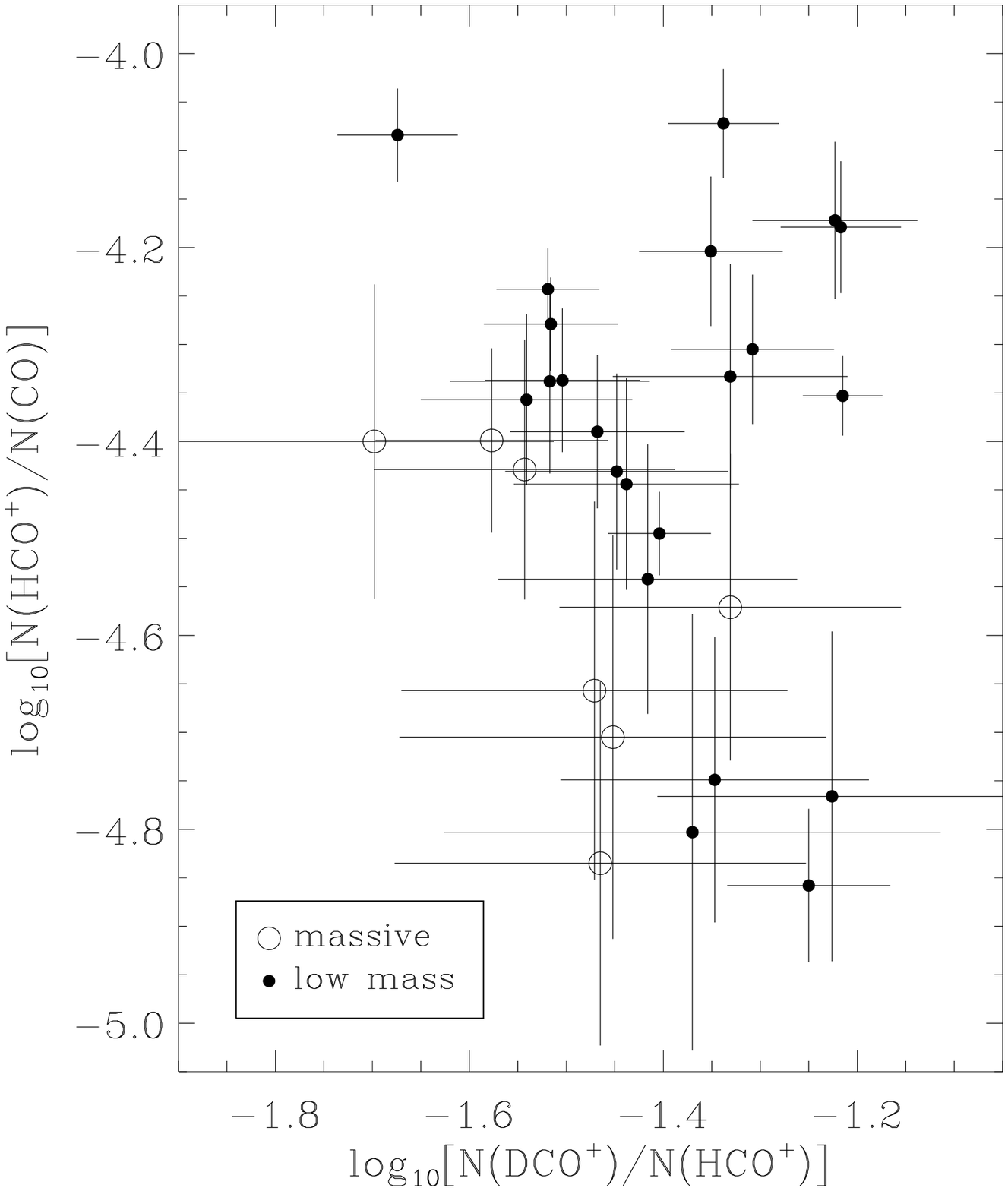}{6in}{0}{70}{60}{-200}{0}
\caption{
Column density ratios for the low mass cores surveyed in Paper I and
the massive cores in the current study. 
}
\label{highlow}
\end{figure}

Figure~\ref{highlow} plots the abundance of \HCOp\ relative to CO against the 
deuterium fractionation ratio: [\DCOp ]/[\HCOp ] for our sample of Orion cores
(as open diamonds).  
Also shown as solid dots are the same ratios for low mass cores taken from
Paper I.  Comparing our data to the low mass cores shows general agreement 
in the same overall range of \HCOp\ abundances.
However, the level of fractionation in deuterium is slightly 
different: the majority of massive cores have very low deuterium 
fractions (log$_{10}$([\DCOp ]/[\HCOp ]) $< -1.45$).  
Moreover, the two cores with the lowest fractionation ratios have outflows that are
traced by CS emission.

\begin{figure}
\figurenum{7}
\plotfiddle{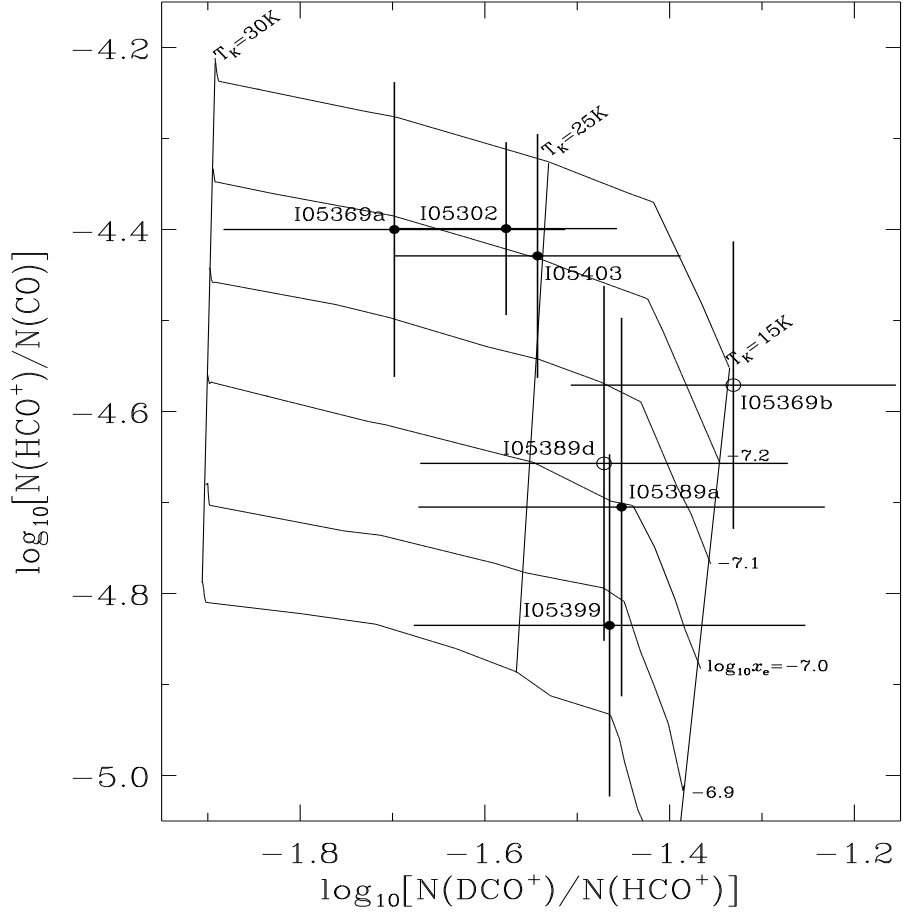}{6in}{0}{70}{60}{-200}{0}
\caption{
Contours of ionization fraction as a function of column density
ratio. 
The logarithm to base 10 of the electron fraction
is labeled in each plot along the right hand side of the contours.
The model is for a density
of n$_{H_2} = 10^{5}$, $\zeta_{H_2} = 5 \times 10^{-17}$ s$^{-1}$, $A_V = 7.5$ mag,
and the normal interstellar radiation field (see Table 6 for listing of other
parameters).
Variations in $y$, the gas kinetic temperature in the model,
is labeled along the upper
side of each plot. Filled circles mark cores associated with stars and open circles
mark the starless cores.
}
\label{model}
\end{figure}

Figure~\ref{model} overlays the best-fit chemical  models on the 
Orion core data (same axes as Figure~\ref{highlow}).  
Vertical lines in Figure~\ref{model}
indicate the model temperature, whereas the horizontal contours denote
the log of electron abundance (log$_{10}x_e$).
This family of models has $\zeta_{H_2} = 5 \times 10^{-17}$, 
$n_{H_2} = 10^{5}$ \cc , carbon and oxygen abundances depleted by 20\% from the
nominal values (see Table 6).
The cores containing stars are plotted as filled
circles and starless sources are open circles. 
As in Paper I the vertical spread in the data is accounted
for by variations in metal ion abundances.  The line for 
log$_{10}(x_e) = -7.3$ in Figure~\ref{model} corresponds to 2 times the
fiducial metal abundances and the line for log$_{10}(x_e) = -6.9$ corresponds
to 10 times the fiducial values.
Thus the range of our observations requires only a small variation in
metal abundances.  As the 
fiducial abundances are already 
heavily depleted from the observed abundances in diffuse clouds (and
even more so from solar values) our results still require highly depleted
metals.
The electron abundances derived for the surveyed Orion cores  
are determined on the basis of the best fit model shown in Figure~\ref{model}.
Values are given in Table~\ref{colden} along with the maximum and minimum abundances
derived from the model and the 1 $\sigma$ observational errors in the ratios.   

There is one major difference
between the models presented here and those constructed for low mass cores (Paper I).
The range of densities observed in low mass cores allowed us to fit models with
constant temperature and a small range of C and O depletions (0.5
to 1.2 times the fiducial values).  However, the densities in the  massive 
cores are constrained to a fairly narrow range, with most
falling around 10$^5$ cm s$^{-1}$ (Table 5).  
At this density
models similar to those presented in Paper I  
are unable to reproduce the low [\DCOp ]/[\HCOp ] ratios observed in
some cores.  For a fixed temperature the lowest ratios require high carbon and oxygen
abundances, to a degree that is inconsistent with observed depletions towards
diffuse clouds.  Moreover the Orion cores observed here are much
denser than diffuse clouds and should, if anything, show larger depletions.

Temperature variations within the sample of massive cores offer a 
simple solution to account for the observed differences.  The main reaction that forms   
\HtwoDp\ (and eventually \DCOp ) is:

\begin{equation}
{\rm H}^+_3 + {\rm HD} \rightleftharpoons {\rm H}_2{\rm D}^+ + {\rm H}_2.
\end{equation}

\noindent The forward reaction (formation of \HtwoDp ) is favored at low temperatures, 
but once the temperature becomes greater than $\sim$20 K the reverse reaction will
begin to contribute significantly and the [\DCOp ]/[\HCOp ] ratio will be lowered. 
Therefore, we account for the 
range of [\DCOp ]/[\HCOp ] ratios in Figure~\ref{model} by varying the
temperature from 15 K to 30 K (vertical lines).  We believe this explanation is plausible because
the cores that contain stars and molecular outflows have the lowest fractionation ratios, whereas the starless
cores have [\DCOp ]/[\HCOp ] ratios close to those observed in the colder low mass cores (Paper I). 
A similar analysis of deuterium fractionation in a wide range of cores by Wootten
et al. (1982) also suggests that temperature variations are necessary to account for the
observed differences.   

Carbon and oxygen depletion is still a variable in the model
calculations.  As demonstrated in Paper I changes in the carbon and
oxygen depletion can alter the electron fraction  
(a depletion of 25\% from the normal values raises $x_e$ by $\sim $25\%).
The best fit to our data requires
an additional depletion of 20\% from the nominal values of [C]/[\Htwo ] = 
$1.46 \times 10^{-4}$ and [O]/[\Htwo ] = $3.52 \times 10^{-4}$.  
However, our data are consistent with C and/or O depletions of 50\% and 
enhancements as high as 125\%, provided that the temperature is still a variable. 
Chemical models of the Orion ridge near the Trapezium cluster, a region that is
quite unique compared to the rest of the Orion cloud, suggest that oxygen atoms
might be more depleted than carbon atoms (giving higher C/O ratios: Bergin et al. 1997a).  
In addition a recent compendium of oxygen depletion measurements
from the Goddard High Resolution Spectrometer suggests that the total oxygen
abundance is homogeneous in the solar vicinity and is two-thirds of the 
solar value (Meyer, Jura, \& Cardelli 1998).  Given the best fit C and O abundances, and 
the range of allowed depletions in the models,
our results do not disagree with the chemical models of 
the Orion ridge or with the recent oxygen measurements.

\section{Discussion}

\subsection{Comparison with Previous Models}

Within the observational and theoretical errors (estimated to be a factor
of 4; see discussion in \S 5.3) 
our electron abundances are consistent with those estimated 
by Wootten et al. (1982).  However, our paper uses a direct comparison to
chemical models, with associated improvements in reaction rates.
Another study observed several molecular ions and neutrals, and with
comparison to current equilibrium chemical models, derived electron abundances
in several massive star forming cores (de Boisanger et al. 1996).  Their chemical
models are similar to ours in 
the sense that the cosmic ray flux is constant for an entire cloud (or complex),
but their sources contain a significantly larger amount of mass and are located
near O and B clusters.
Our approach of fixing the ionization rate
is different from that found in Caselli et
al. (1998) who derived electron abundances for a sample of low mass cores (mostly
in Taurus and Ophiuchus).
Caselli et al. (1998) used the cosmic ray ionization rate as a true variable parameter in
their analysis and account for the spread in [\DCOp ]/[\HCOp ] and 
[\HCOp ]/[CO] ratios 
through an intrinsic variation of the cosmic ray flux from core to core.
Our analysis takes the flux of cosmic rays as a variable only to fit the entire sample.
Slight differences in temperature between cores can 
account for the changes in deuterium fractionation between the different
cores.    
Since our small sample of cores is located solely in the Orion complex,
it is likely that the exteriors of these cores are bathed in a similar cosmic ray flux. 
A single value for the cosmic ray flux for a region several parsecs in size is
also supported by the large-scale consistency of the diffuse gamma-ray emission, which is
believed to originate from cosmic ray interactions with matter, 
with the large-scale distribution of the ISM (see Bertsch et al. 1993 and references
therein).

An important source of concern in our models is the requirement that metal ions vary in
abundance among our small sample of cores by a factor of 6.  As noted in \S 4.2 
this small variation still implies that metal abundances are heavily depleted from
diffuse cloud values.  Such a variation could be found if these cores have
slightly different density structures and formation histories.  In this regard
it is encouraging that the electron abundances and predicted metal depletions in 
the two sources with multiple cores are quite similar. 
Another complication is that the non-linear character of chemical equations
has been shown to give rise to multiple solutions with high- and low-ionization
phases (Le Bourlot, Pineau des For\^ets, \& Roueff 1995).  
These solutions are found to be dependent on the initial conditions.
Thus, the fractional ionization may not be uniquely determined from the molecular
column density ratios used here.  However, in our models we have examined a
large fraction of the parameter space (abundances of metals, cosmic ray ionization
rate, temperature, and density).  In all, over 1200 models were run for this
paper and in Paper I to find the the best match to the data.  In no case was a bistable
solution found and our results are consistent with the low-ionization phase 
(see Plume et al. 1998). 
Finally, the
non-inclusion of gas-grain interactions could be important.   
For a detailed discussion of these
effects and other possible sources of electrons the reader is referred to the
discussion in Paper I.

\subsection{The Effect of the Radiation Field}

In terms of electron abundances and the radiation field there are two possible effects.
One concern is that the local radiation from an associated young star may heat
the surrounding molecular envelope and produce a gradient in the deuterium fractionation 
along the line of sight.   In contrast the \HCOp\ abundance will not be as strongly
affected.
Using Figure~\ref{model} we can attempt to quantify these effects. 
Along a line of constant [\HCOp ]/[CO], the derived electron fraction increases
by a factor of 1.6 as the deuterium fractionation ratio declines.  
Thus, a gradient in D-fractionation will result in an overestimation of 
the actual electron abundance.

For most cases we have avoided this particular 
concern by performing our analysis along lines of sight away from the stellar
position.  The analysis for one source, IRAS05389--0756a, is performed
near the position of the IRAS source and it is possible that this 
could affect our determination of the electron abundance.  If we examine
Figure~\ref{model} the [\DCOp ]/[\HCOp ] ratio for IRAS05389 core (a), which is
near the IR source, and core (d), which is apparently unassociated with star formation,
are nearly identical.  This suggests that, for this source at least, 
such line of sight effects may not be a dominating factor. 
To probe this question further we
can compare the [\DCOp ]/[\HCOp ] ratio computed at the stellar position with
a single position away from the source for all surveyed cores.  
We find that, on average, the change is
20\% in the direction of lower fractionation.  A typical statistical error in the D/H 
ratio is $\sim 30$\% which suggests that, on average, 
this effect is not important at the observed 
spatial resolution.
One source, IRAS05302--0537, one of the more luminous stars 
in our sample ($L_{IR} = 57 L_{\odot}$), shows a
67\% decrease.   Thus, there is some (inconclusive and limited) 
evidence of this effect.  A larger sample of sources and
data with greater spatial resolution and sensitivity will be
required to examine this question in more detail.
 
A second potential effect is from the external radiation field.
McKee (1989) 
showed that for a homogeneous gas layer illuminated by the normal
interstellar radiation field, cosmic ray ionization and photoionization
contributed equally at $A_V \sim 4$.  We have investigated the effects of the 
radiation field on our chemical calculations 
by lowering the visual extinction or, equivalently, raising the external UV field.
Since these cores are located in the Orion cloud, which
have several local sources of intense radiation (e.g. Genzel \& Stutzki 1989), we
have investigated raising the UV radiation field by a factor of 500.
We find that
our results are unaltered.  Raising the UV field to  
$\chi = 500$ corresponds to lowering the visual extinction to $A_V \sim 2 - 3$ (depending
on the molecular photoabsorption cross-section), and at these depths CO molecules
are still partially shielded from radiation.
It is important to note that our models do account for
the self-shielding of CO. 
 
We have further investigated the effect of the radiation field by
comparing the [DCO$^+$]/[HCO$^+$] ratio and the H$^{13}$CO$^+$
abundances in the core edges to those found in the center.
The separation between the core edges and center is defined as the
contour at which the DCO$^+$ J = $1\rightarrow0$ integrated intensity
is equal to half the peak value. For a particular molecule,
all spectra that lay in the ``edge''
region
($\int{T_RdV(DCO^+)} < 1/2\int{T_RdV_{peak}}$)
were combined to produce a single ``edge'' spectrum whereas those that fell in
the ``center'' were combined to produce a single ``center'' spectrum.
In this fashion, we obtained average ``edge'' and ``center'' spectra for
each molecule in each of the cores in our sample.
This analysis was performed on IRAS05399--0121 and
IRAS05302--0537, since these cores are centrally condensed and have
enough spectra to construct meaningful averages.
We find that, in IRAS05399--0121, the [DCO$^+$]/[HCO$^+$] ratio
increases by only 0.1 in the log between the edge to the center, and
the [HCO$^+$]/[CO] ratio increases by less than 0.3 in the log.
In IRAS05302--0537, both the [DCO$^+$]/[HCO$^+$] and [HCO$^+$]/[CO] ratios
increase by only 0.1 in the log as one goes from the edge to the center.
Inspection of Figure~\ref{model} shows that these small changes in the abundance
ratios lie well within our observational error bars and have a minimal
effect upon the derived electron abundance.  
Thus, we find no difference in the deuterium fraction or the \HthCOp\ abundance
between the low visual extinction core edges and the core centers.
The ionization fraction implied by this result depends on the unknown density 
structure of these cores.  For a constant density (and temperature) our models will 
predict similar ionization levels between edge and center of the maps.  
If these cores have a power-law density profile as found for some low mass cores 
(Goldsmith \& Arquilla 1985; Ward-Thompson et al. 1994) and suggested by Zavagno 
et al. (1997) for IRAS05389--0756, then the model predicts a 
slightly higher level of ionization at the core edges. 
However, for both scenarios cosmic ray ionization
provides the main source of electrons.

In order to more carefully investigate the effects of photoionization 
we can utilize the model
of Jansen et al. (1995) who examined the depth dependent chemistry of IC63, which
has a B star located $\sim 1$ pc away.  
This work 
includes a detailed 
treatment of the observed UV field of the B star, and its coupling to the
photodissociation rates for various molecules.
In their model, the abundance of \HCOp\ is $\sim 10^{-11}$ near the core edges, 
due to the increased abundance of photo-produced electrons.   At larger depths
($A_V >$ 4) the abundance increases to $\sim 10^{-9}$.   Since our results 
show no evidence for a decline in  the \HCOp\ abundance we conclude that we are
tracing gas that is shielded from  UV radiation and ionized primarily by cosmic rays.
The lack of abundance variation from the edge to the center
is similar to recent measurements of abundances across the M17 molecular cloud/ionization
front interface where 
molecular abundances show little variation despite large changes in the radiation field
(Bergin et al. 1997b).   

\subsection{Electron Abundances and Star Formation}

It is worthwhile to compare the electron abundances between our sample of
massive and low mass objects.  
In order to extend our study towards even more massive cores 
we use previous studies in the literature.  Wootten et al. (1982)
surveyed many sources for emission from  DCO$^{+}$, \HthCOp , and \thCO .  We have
used their raw data from NGC1333 and in Serpens (Ser MC1 in their notation) to derive 
column densities and, with comparison to our models, electron abundances.  We 
find that $x_e = 4 \times 10^{-8}$ in NGC1333 and $x_e = 5 \times 10^{-8}$ in Serpens.
We can also draw upon the recent work of de Boisanger et al. (1996) who
examined electron abundances in denser, very 
massive sources: W3 IRS5 ($x_e = 1 \times 10^{-8}$) and 
NGC2264 ($x_e = 5 \times 10^{-9}$).   Finally, we compare the observations 
of Bergin et al. (1997a) 
in the OMC-1 core to our chemical models
to obtain
$x_e = 2 \times 10^{-8}$.  

Before we compare the electron abundances between our work here and in Paper I
with those of de Boisanger et al. (1996) it is worthwhile to estimate the 
total uncertainty in the derived electron fraction.  In Table~7 we provide an estimate
of the minimum and maximum electron abundance that is consistent with the 
statistical errors in the observed column density ratios.  The average range is
$\Delta$log$_{10}$($x_e) = 0.3$, which suggests that the uncertainty
is, roughly, a factor of $\sim$1.4.   This range does not include 
any systematic error that might be introduced by our assumptions with regard
to the physical parameters such as the density and temperature.  For the massive
cores sampled here, the uncertainty in the determined density, including statistical
and systematic effects, is typically a factor of $\leq$3 (see Table 5).  This 
produces a factor of two uncertainty in the derived ionization
fraction.  For the temperature, if we assume that the allowed range is between 
15 and 25 K then this is an additional factor of 1.5.  Therefore, with the additional
assumption that the cosmic ray ionization rate is the same for each core 
(see \S 5.1), the total uncertainty
in the derived electron fraction is a factor of $\sim$4.
The same reasoning applies to the analysis in Paper I, while
de Boisanger et al. (1996) find that their chemical model predictions are within a 
factor of two of the observed abundances of 7 ions.   

Perhaps the most meaningful comparison between electron abundances among the 
overall sample of cores would compare the level of ionization
to the density of each core.  However, there is no comprehensive 
examination of the density using similar models, observations, and methods 
for the entire range of objects.  The densities of the very massive cores
(e.g. OMC-1)
have been determined via excitation analyses of CS, \HtwoCO , and \HCthreeN ,
with values $_{\sim}^{>} 10^6$ \cc\ (e.g. Mundy et al. 1985; Plume et al. 1997; 
Bergin, Snell, \& Goldsmith 1997).  The massive cores 
studied in this work have lower densities, $n_{H_2} \sim 10^5$ \cc , from
our CS analysis.  However, the low mass cores have not been systematically
studied using these techniques, although Caselli \& Myers
(1995) show that for a given radius low mass cores have 
smaller densities by about a factor of three (i.e. massive cores
are larger and denser).  
 
\begin{figure}
\figurenum{8}
\plotfiddle{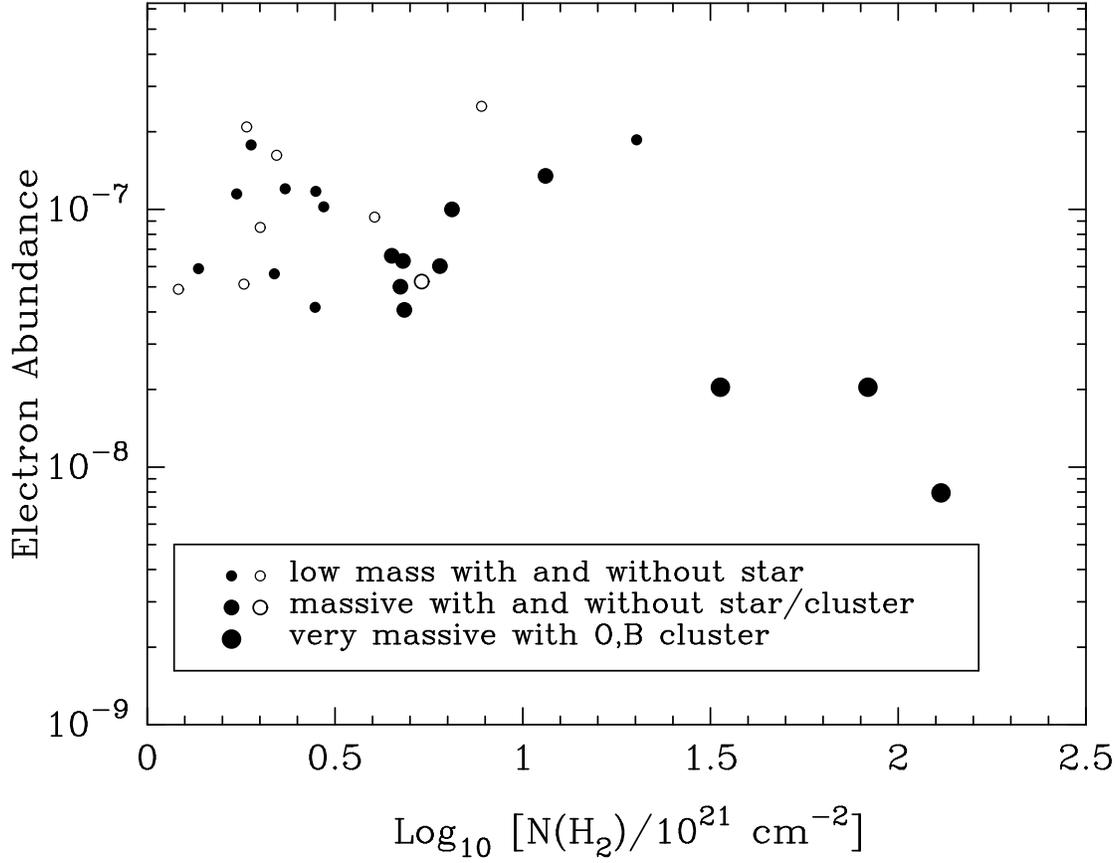}{5in}{0}{90}{90}{-320}{-270}
\caption{
Plot of the electron abundance shown as a function of total H$_2$ column density.
This plot combines electron abundances derived in this work, from Paper I,
from de Boisanger et al. (1996), and using the data from Wootten et al. (1982) and
Bergin et al. (1997a).  
}
\label{xemvir}
\end{figure}

Since the density is not consistently derived for each
source we have opted to use the total \Htwo\ column density along the
line of sight.   The \Htwo\ column density is determined using N(C$^{18}$O), 
and an assumed fractional abundance of $5 \times 10^{-7}$.
\Htwo\ column densities for W3 IRS5 and NGC2264
are taken from de Boisanger et al. (1996), while for OMC-1 we use the \CeiO\
column densities in Bergin et al. (1997a).
The electron abundance shown as a function of N(\Htwo ) is presented in
Figure~\ref{xemvir}. 
Examining Figure~\ref{xemvir}, the abundances for 
the low mass cores show a large degree of scatter, with a median value of 
$< x_e > \sim 10^{-7}$.   The massive cores (along with
Serpens and NGC1333) overlap with the low mass sources, but are more clustered with 
a lower median electron abundance of 
6 $\times 10^{-8}$.   Given the estimated uncertainties, which will be reduced by
$\sqrt{N}$ (N $=$ the number of cores) for the median, and the small number of
data points for massive cores, the
differences between these two samples are only suggestive.
However, there is a significant difference
in the ionization fraction between the very massive (and densest) cores 
(W3, OMC-1, and NGC2264) and the 
low mass cores.     
These results suggest that the regions with the highest column and volume densities 
also have lower electron abundances, as would
be expected from the conventional picture of cosmic ray ionization.
However, these results must be viewed with caution as there are only 3 data points for
very massive sources and the electron abundances for these sources
were derived from a different (although equally valid) model. 
It is noteworthy that there appears to be
no difference in the ionization fraction 
between cores directly associated with star formation and
cores without any activity.

\subsection{Ion-Neutral Coupling in Massive Cores}

In Paper I we examined the question of the ion-neutral coupling in
low mass cores in terms of the wave coupling parameter.  
The wave coupling parameter, $W$, is defined
as the ratio of the core size, $R$, to cutoff wavelength $\lambda_0$. The 
cutoff wavelength is defined by $\lambda_0 = \pi v_A/[n_i <\sigma_{in} v>]$, where
$v_A$ is the Alfv\'en speed, $n_i$ the ion density, and $<\sigma_{in} v>$ the ion-neutral
collision coefficient.  If W $\gg 1$ the field and gas are coupled over a 
large range of size scales.   For W $\ll 1$ the
ions are decoupled from the field and MHD waves are suppressed (see Paper I for a 
more comprehensive discussion).  

The ionization fraction is determined by the balance of cosmic ray ionization and
recombination, along with any contribution from metal ions with low ionization potential
and longer recombination timescales.  The relation between these values can
be expressed as

\begin{equation}
x_e = C_i n_{H_2}^{-1/2}\zeta_{H_2}^{1/2}
\end{equation}

\noindent where $C_i$ is a constant that contains the relative contributions of molecular
ions and metals to the ionization balance.  
McKee (1989) 
derives $C_i = 3.2 \times 10^3$ cm$^{-3/2}$s$^{1/2}$ for an idealized model of cosmic ray 
ionization. In Paper I we derived a value of $C_i = 2.0 
\times 10^3$ cm$^{-3/2}$s$^{1/2}$.  If we use
our average electron abundance of $8 \times
10^{-8}$ and density of 10$^{5}$ \cc , we find, for the massive cores in Orion,
a value of $C_i = 3.6 \times
10^3$ cm$^{-3/2}$s$^{1/2}$.   
Considering the error in the electron
fraction the difference between the value derived here
and for low mass cores is not significant.

For cores which are ionized primarily by cosmic
rays and are dominated by non-thermal motions,
the wave coupling parameter 
is independent of density and can be expressed as:

\begin{equation}
W = \frac{C_i <\sigma_i v>}{\pi} \left[\frac{10 \zeta_{H_2}}{4\pi mG}\right]^{1/2}
\end{equation}

\noindent where $<\sigma_i v> = 1.5 \times 10^{-9}$ cm$^3$s$^{-1}$, 
and $m$ is the mean molecular mass per particle.  The equation is listed
for the case of maximum turbulence as appropriate for massive cores.   
Both of these conditions, high non-thermal width
and cosmic ray ionization, are satisfied for the massive cores in our study and
using our best fit value of the ionization rate,
$\zeta_{H_2} = 5 \times 10^{-17}$ s$^{-1}$, gives $W = 20$.  

        Thus the field-neutral coupling W of the seven massive cores
analyzed here is similar to, or perhaps slightly larger than, that deduced
for the 20 low mass cores in Paper I (W = 11).  For both samples, the coupling is
``marginal,'' i.e. strong enough to allow propagation of MHD waves, but weak
enough for some wave damping to occur (Martin, Heyvaerts, \& Priest 1997,
Nakano 1998, Myers \& Lazarian 1998).  This suggests that both low mass and
massive cores are more vulnerable to fragmentation and dissipation of their
turbulence than are their lower-density, better-coupled environs (Myers
1997).

        However, the massive cores considered in this paper appear better
able to form stellar groups and clusters than the low mass cores in Paper
I, for two reasons.  First, the massive cores can be ``clumpier'' than 
low mass cores, because they have a greater ratio of turbulent to
thermal speeds than do the low mass cores (Myers 1998).   Second, the
typical massive core is a factor of $\sim$2 larger than the typical low mass
core, so the massive core has more high-density gas and more volume
available in which to form stars, than does the typical low mass core.
It will be useful to extend the observations and analysis presented
here to more ``very'' massive regions forming OB stars and rich clusters.

\section{Conclusions}

We have mapped the emission of three molecular ions (\DCOp , \HthCOp , and \NtwoHp )
and two neutral species (CS and \CeiO) in massive  dense cores.
In all we survey 5 cores that are currently associated with star formation and 2
that are starless.  These cores are located in the L1630 and L1641 star forming regions, 
and are more massive, turbulent, and larger than the sample of low mass cores subject
to a similar study in Paper I. 
We have used these observations to derive electron 
abundances using the same method and chemical model as applied to 
the low mass cores in Paper I.  The principal results
of this study are:

1) With use of observations to constrain variables in the chemical calculations,
we have constrained the electron fraction in our sample to lie within
a small range --6.9 $<$ log$_{10}$($x_e$) $<$ --7.3, with an average value of --7.11 $\pm 0.15$.
These values are consistent with previous analyses of ionization fractions in
other star forming regions.   No difference is found between the electron abundances
inferred for cores with and without stars.
However, the cores directly 
associated with stars and molecular outflows have lower levels of deuterium 
fractionation, which can be attributed to differences in temperature between the 
various sources.

2) Among our sample of cores we have searched for differences in the ionization
level between the core edges and center.  For the two cores that we examined
in detail, the observed
levels of deuterium 
fractionation and \HCOp\ abundances do not change with position.  The exact
value of ionization fraction implied by the constant ratios depends quite critically
on the unknown density/temperature structure of these sources.  
However, the lack of variation
suggests that even the core edges probed by these high dipole moment molecules are 
shielded from ultra-violet radiation and cosmic rays remain the primary source of
ionization.  

3) Using the  previously determined electron fractions in low mass cores
forming single isolated stars, and other values for very massive cores associated with
O and B stars, we have systematically examined electron abundances
over a wide spectrum of star formation activity.  The very 
massive sources, such as OMC-1 or W3, have ionization fractions typically an order
of magnitude below those estimated for low mass cores where $< x_e > \sim 10^{-7}$.  

4) The issue of ion-neutral coupling is examined 
as a function of core properties.
We find that the size of the decoupled region is larger for 
more massive cores; thus massive
cores have a greater amount of mass with lower field-gas coupling
available to form stars.

We are grateful to K. Strom and L. Allen for making available their near-infrared
data in L1641.  We also thank P. Caselli and M. Walmsley
for several discussions on the topic of
electron abundances and A. Dalgarno for discussions on cosmic ray ionization.
E.A.B. and R.P. thank the AAS for a  small research grant to support this work and also
acknowledge support from NASA's SWAS contract to the Smithsonian Institution (NAS5-30702).
P.C.M. and J.P.W. acknowledge partial support from NASA Origins of Solar Systems Program
grant NAGW-374.

\newpage
 
\begin{deluxetable}{lcc}
\tablenum{1}
\tablewidth{4.5in}
\tablecolumns{3}
\tablecaption{Map Central Positions}
\tablehead{
\colhead{Source} &
\colhead{$\alpha(1950)$} &
\colhead{$\delta(1950)$} \\ 
}
\startdata
IRAS05399--0121a & 05$^h$39$^m$55.1$^s$ & --01$^{\circ}$21$'$24$''$ \nl
IRAS05302--0537a & 05 30 15.8 & --05 37 32  \nl
IRAS05369--0728b & 05 36 51.1 & --07 25 34  \nl
IRAS05389--0756a & 05 38 56.6 & --07 56 40  \nl
IRAS05403--0818a  & 05 40 23.3 & --08 18 26  \nl
\enddata
\label{centralpos}
\end{deluxetable}

\newpage
\begin{deluxetable}{lllcccc}
\tablenum{2}
\tablecolumns{7}
\tablecaption{Observed Transitions and Telescope Parameters}
\tablehead{
\colhead{Molecule} &
\colhead{Transition } &
\colhead{$\nu$(GHz)} &
\colhead{$E_u$ (K)} &
\colhead{Telescope} & 
\colhead{$\theta_{MB}$} &
\colhead{$\eta_{MB}$} \\ 
}
\startdata
DCO$^{+}$\tablenotemark{a}   & $J = 1 \rightarrow 0$ & 72.039331 & 3.46  &  NRAO  & 80$''$  & 0.93 \nl
H$^{13}$CO$^{+}$\tablenotemark{a} & $J = 1 \rightarrow 0$ & 86.754329 & 4.16 & NRAO & 67$''$ & 0.91 \nl
N$_2$H$^+$ & $J = 1 \rightarrow 0$ & 93.1737\tablenotemark{b} & 4.47 & FCRAO & 56$''$ & 0.51 \nl
CS & $J = 2 \rightarrow 1$  & 97.981011 & 2.35 & FCRAO & 53$''$ & 0.50 \nl
C$^{18}$O & $J = 1 \rightarrow 0$ & 109.782182 & 5.27 & FCRAO & 47$''$ & 0.46 \nl
CS & $J = 3 \rightarrow 2$ & 146.969049 & 7.05 & NRAO & 43$''$ & 0.75 \nl
\enddata
\tablenotetext{a}{NRAO efficiencies listed are $\eta_M^*$.}
\tablenotetext{b}{Line off center to obtain all hyperfine components in spectrum.}
\end{deluxetable}

\newpage

\begin{table}
\tablenum{3}
\label{line_param}
\plotfiddle{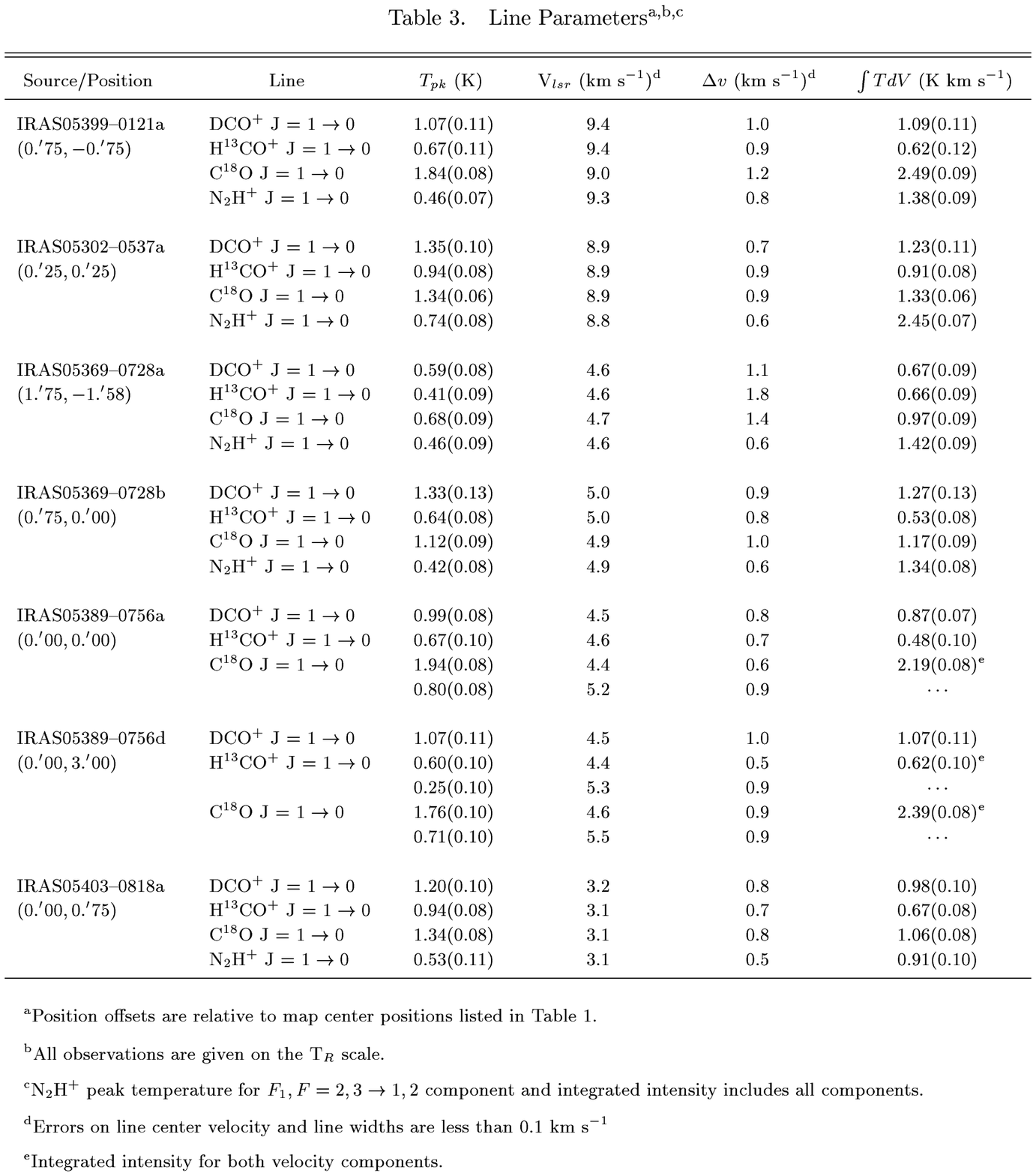}{8in}{0}{100}{100}{-300}{-50}
\end{table}

\newpage
\begin{deluxetable}{lcccccc}
\small
\tablenum{4}
\tablewidth{7.0in}
\tablecolumns{7}
\tablecaption{CS Line Parameters\tablenotemark{a,b}}
\tablehead{
\multicolumn{1}{c}{Source} &
\multicolumn{3}{c}{J $= 2 \rightarrow 1$} &
\multicolumn{3}{c}{J $= 3 \rightarrow 2$} \\ 
\colhead{} &
\colhead{T$_{R}$ (K)} &
\colhead{V$_{lsr}$ (km s$^{-1}$)} &
\colhead{$\Delta v$ (km s$^{-1}$)} &
\colhead{T$_{R}$ (K)} &
\colhead{V$_{lsr}$ (km s$^{-1}$)} &
\colhead{$\Delta v$ (km s$^{-1}$)} 
}
\startdata
IRAS05399--0121a   & 1.08(0.10) & 9.0  & 1.1 &1.32(0.06) & 9.0 &1.2 \nl
IRAS05302--0537a  & 0.82(0.10) & 8.5  & 2.3 &1.38(0.06) & 8.3 &1.8 \nl
IRAS05369--0728a  & 1.00(0.10) & 4.3  & 0.8 &1.17(0.07) & 4.6 &3.5 \nl
IRAS05369--0728b  & 0.92(0.07) & 5.0  & 1.5 &0.91(0.06) & 4.9 &1.4 \nl
IRAS05389--0756a  & 0.56(0.09) & 4.7  & 1.4 &0.42(0.06) & 4.6 &1.2 \nl
IRAS05403--0818a   & 1.09(0.14) & 3.2  & 0.7 &1.14(0.06) & 3.1 &1.2 \nl
\enddata
\tablenotetext{a}{Line parameters determined from observations of central positions listed in Table 1, except IRAS05369-0728a which is at ($1'.75, -1'.58$) relative to core (b)}
\tablenotetext{b}{Errors for line center velocity and line width are less than 0.1 km s$^{-1}$}
\label{CSline_param}
\end{deluxetable}

\newpage
\begin{deluxetable}{lcccc}
\small
\tablenum{5}
\tablecolumns{5}
\tablecaption{Core Densities and CS Column Densities\tablenotemark{a,b}}
\tablehead{
\multicolumn{1}{c}{} &
\multicolumn{2}{c}{T$_k$ = 15 K} &
\multicolumn{2}{c}{T$_k$ = 25 K} \\
\colhead{Source} &
\colhead{log$_{10}$ n$_{H_2}$ } &
\colhead{log$_{10}$ N(CS)} &
\colhead{log$_{10}$ n$_{H_2}$} &
\colhead{log$_{10}$ N(CS)} 
}
\startdata
IRAS05399--0121a   &  5.2(0.3) & 13.2(0.2) & 5.0(0.3) & 13.2(0.3) \nl
IRAS05302--0537a   &  5.6(0.4) & 13.2(0.3) & 5.4(0.3) & 13.1(0.3) \nl
IRAS05369--0728a  & 5.2(0.4) & 13.4(0.5) & 4.9(0.4) & 13.4(0.6) \nl
IRAS05369--0728b  & 4.9(0.5) & 13.3(0.3) & 4.7(0.6) & 13.2(0.4) \nl
IRAS05389--0756a  &  4.5(1.2) & 13.3(1.5) & 4.1(2.0) & 13.2(0.3) \nl
IRAS05403--0818a   & 5.0(0.4) & 13.2(0.3) & 4.7(0.5) & 13.2(0.4) \nl
\enddata
\tablenotetext{a}{Densities are in units of cm$^{-3}$ and column densities in cm$^{-2}$}
\tablenotetext{b}{Values derived towards central positions listed in Table 1, except
IRAS05369-0728a which is at ($1'.75, -1'.58$) relative to core (b)}
\label{CSden}
\end{deluxetable}
 
\newpage
\begin{deluxetable}{ll}
\tablenum{6}
\tablecolumns{2}
\tablewidth{3.0in}
\tablecaption{Adopted Model Parameters\tablenotemark{a}}
\tablehead{
\colhead{Parameter} &
\colhead{Value} 
}
\startdata
$\zeta_{H_2}$ & 5 $\times 10^{-17}$ s$^{-1}$  \nl
$n_{H_2}$ & 1.0 $\times 10^5$ cm$^{-3}$  \nl
$T_k$ & $15 - 30$ K  \nl
$A_V$ & 7.5 mag  \nl
$\chi$ & 1.0  \nl
He & 0.28  \nl
HD & 2.8 $\times 10^{-4}$  \nl
C$^{+}$\tablenotemark{b} & 1.17 $\times 10^{-4}$  \nl
O\tablenotemark{b} & 2.82 $\times 10^{-4}$  \nl
N & 4.28 $\times 10^{-5}$  \nl
M$^{+}$\tablenotemark{c} & 2 -- 10 M$^+_0$  \nl
\enddata
\tablenotetext{a}{Listed abundances are relative to H$_2$}
\tablenotetext{b}{C and O abundances are 80\% of fiducial values listed in Table 4 of Paper I -- and are $\sim$20\% of the solar abundances.}
\tablenotetext{c}{M$^+$ = S$^+$ $+$ Si$^+$ $+$ Fe$^+$ $+$ Mg$^+$ + P$^+$ and M$^+_0$ is the
fiducial abundances of these species listed in Table 4 of Paper I .}
\label{tab_model}
\end{deluxetable}
 
\newpage
\begin{deluxetable}{lccccccc}
\footnotesize
\tablenum{7}
\tablewidth{7.5in}
\tablecolumns{8}
\tablecaption{Molecular Column Densities and Electron Abundances\tablenotemark{a,b}}
\tablehead{
\colhead{Source} &
\colhead{log$_{10}$ N(DCO$^{+}$)} &
\colhead{log$_{10}$ N(H$^{13}$CO$^{+}$)} &
\colhead{log$_{10}$ N(N$_{2}$H$^{+}$)} &
\colhead{log$_{10}$ N(C$^{18}$O)} &
\colhead{$x_e$} &
\colhead{Min} &
\colhead{Max} \\
}
\startdata
IRAS05399--0121a  &12.15(11.2) & 11.84(11.1) & 12.60(11.4) &15.76(14.3)&--6.87&--6.74&--7.03\nl
IRAS05302--0537a  &12.21(11.1) & 12.01(10.9) & 12.84(11.3) &15.48(14.2)&--7.22&--7.14&--7.31\nl
IRAS05369--0728a &11.95(11.1) & 11.87(11.0) & 12.61(11.4) &15.35(14.3)&--7.18&--7.05&--7.33\nl
IRAS05369--0728b &12.22(11.2) & 11.78(11.1) & 12.59(11.3) &15.43(14.3)&--7.28&--7.15&--7.40\nl
IRAS05389--0756a &12.06(11.0) & 11.73(11.0) & \nodata     &15.51(14.1)&--7.00&--6.83&--7.17\nl
IRAS05389--0756d &12.15(11.1) & 11.84(11.1) & \nodata     &15.58(14.5)&--7.04&--6.86&--7.20\nl
IRAS05403--0818a  &12.11(11.1) & 11.87(10.9) & 12.42(11.5) &15.38(14.2)&--7.20&--7.08&--7.31\nl
\enddata
\tablenotetext{a}{Column densities in units of cm$^{-2}$.}
\tablenotetext{b}{Electron abundances are relative to H$_2$.}
\label{colden}
\end{deluxetable}

\clearpage

\newpage
\begin{thebibliography}{}
 
\bibitem[Allen 1996]{Allen96} Allen, L.E. 1996, Ph.D. Thesis, University of Massachusetts

\bibitem[Anglada et al. 1989]{Angladaetal89} Anglada, G., Rodriguez, L.,
Estalella, R., Torrelles, J.M., Ho, P.T.P., Cant\'{o}, J., L\'opez, R.,
\& Verdes-Montenegro, L. 1989, ApJ, 341, 208

\bibitem[Anglada et al. 1992]{Angladaetal92} Anglada, G., Rodriguez, L.F., Cant\'{o},
J., Estalella, R., Torrelles, J.M. 1992, ApJ, 395, 494

\bibitem[Bally, Langer, \& Liu 1991]{BLL91} Bally, J., Langer, W.D., \& Liu, W. 1991, ApJ,
393, 645

\bibitem[Benson \& Myers 1989]{BM89} Benson, P.J. \& Myers, P.C. 1989, ApJS, 71, 89

\bibitem[Bergin \& Langer 1997]{BL97} Bergin, E.A., \& Langer, W.D. 1997, ApJ, 486, 316

\bibitem[Bergin et al. 1997a]{BGSL97} Bergin, E. A., Goldsmith, P. F., Snell, R.L., \& Langer, W.D. 1997a, ApJ, 482, 285

\bibitem[Bergin et al. 1997b]{BUGSIS97} Bergin, E.A., Ungerechts, H., Goldsmith, P.F., Snell, R.L., Irvine, W.M., Schloerb, F.P. 1997b, ApJ, 482, 267

\bibitem[Bergin, Snell, \& Goldsmith 1996]{BSG96} Bergin, E.A., Snell, R.L., \& Goldsmith, P.F. 1996, ApJ, 460, 343

\bibitem[Bertsch et al. 1993]{Bertschetal93} Bertsch, D. L., Dame, T. M., Fichtel, C. E., Hunter, S. D., Sreekumar, P., Stacy, J. G., \& Thaddeus, P. 1993, ApJ, 416, 587

\bibitem[de Boisanger et al. 1996]{dBHvD96} de Boisanger, C., Helmich, F.P., \&
van Dishoeck, E.F. 1996, A\&A, 310, 315

\bibitem[Caselli et al. 1998]{CWTH98} Caselli, P., Walmsley, C.M., Terzieva, R., \&
Herbst, E. 1997, A\&A, in press 

\bibitem[Caselli \& Myers 1995]{CM95} Caselli, P. \& Myers, P.C., 1995, ApJ, 446, 665

\bibitem[Fukui et al. 1986]{Fukuietal86} Fukui, Y., Sugitani, K., Takaba, H., Iwata, T., Mizuno, A., Ogawa, H., Kawabata, K. 1986, ApJ, 811, 85L

\bibitem[Genzel \& Stutzki 1989]{GS89} Genzel, R. \& Stutzki, J. 1989, ARA\&A, 27, 41


\bibitem[Goldsmith 1987]{G87}
Goldsmith, P.F. 1987, in Interstellar Processes,
ed. D.J. Hollenbach
\& H.A. Thronson (Dordrecht: Reidel), 51 

\bibitem[Goldsmith \& Arquilla 1985]{GA85}
Goldsmith, P.F. \& Arquilla, R. 1985, ApJ, 297, 436

\bibitem[Goldsmith \& Langer 1978]{GL78}
Goldsmith, P.F. \& Langer, W.D. 1978, ApJ, 222, 881

\bibitem[Gu\'{e}lin, Langer,  \& Wilson 1982]{GLW92} Gu\'{e}lin, M., Langer, W.D., \& Wilson, R.W. 1982, \aap, 107, 107

\bibitem[Habing 1968]{Habing68} Habing, H. J. 1968, Bull. Astron. Inst. Netherlands, 19 421

\bibitem[Harju, Walmsley, \& Wouterloot 1993]{HWW93} Harju, J., Walmsley, C.M., \& Wouterloot, J.G.A. 1993, A\&ASS, 98, 51 (HWW)

\bibitem[Jansen et al. 1995]{JvDBSS95} Jansen, D.J., van Dishoeck, E.F., Black, J.H., Spaans, M., \& Sosin, C. 1995, A\&A, 302, 223

\bibitem[Lacy et al. 1994]{LKGT94} Lacy, J.H., Knacke, R., Geballe, T.R., \& Tokunaga, A.T.
1994, ApJ, 428, 69L

\bibitem[Lada, Evans, \& Falgarone 1997]{LEF97} Lada, E.A., Evans, N.J. II, \& Falgarone, E. 1997, ApJ, 488, 286

\bibitem[Lada \& Lada 1991]{LL91} Lada, C.J. \& Lada, E.A. 1991, in The Formation and Evolution of Stellar Clusters, ed. K. Janes (San Francisco: ASP Conf. Ser. 13), 3

\bibitem[Lada et al. 1991]{LDEG91} Lada, E.A., DePoy, D.L., Evans, N.J. II, \& Gatley,
I. 1991a, ApJ, 371, 171

\bibitem[Lada, Bally, \& Stark 1991]{LBS91} Lada, E.A., Bally, J., \& Stark, A. 1991b,
ApJ, 368, 432 

\bibitem[Launhardt et al. 1996]{Launhardt_etal96} Launhardt, R., Mezger, P.G., Haslam,
C.G.T., Kreysa, E., Lemke, R., Sievers, A., \& Zylka, R. 1996, A\&A, 312, 569.

\bibitem[Le Bourlot, Pineau des For\^ets, \& Roueff 1995]{LPR95} Le Bourlot, J.,
Pineau des For\^ets, G., \& Roueff, E. (1995), A\&A, 297, 251

\bibitem[Li, Evans, \& Lada 1997]{LEL97} Li, W., Evans, N.J. II, \& Lada, E.A. 1997, ApJ, 488, 277

\bibitem[Maddalena et al. 1986]{MMMT86} Maddalena, R.J., Morris, M., Moscowitz, J., \&
Thaddeus, P. 1986, ApJ, 303, 375

\bibitem[Martin, Heyvaerts, \& Priest 1997]{MHP97} Martin, C. E., Heyvaerts, J., \& Priest, E. R. 1997, A\&A, 326, 1176

\bibitem[McKee et al. 1993]{MZGH93} McKee, C. F., Zweibel, E.G., Goodman, A.A., \&
Heiles, C. 1993, in Protostars
and Planets III, eds. E. H. Levy, J. I. Lunine, and M. S. Matthews (Tucson:
University of Arizona Press), 327

\bibitem[McKee 1989]{McKee89} McKee, C. F., 1989, ApJ, 345, 782

\bibitem[Meehan et al. 1998]{MWCMW98} Meehan, L.S.G., Wilking, B.A., Claussen, M.J.,
Mundy, L.G., \& Wootten, A. 1998, ApJ, 115, 1599

\bibitem[Meyer, Jura, \& Cardelli 1998]{MJC98} Meyer, D. M., Jura, M., Cardelli, J. A.
1998, ApJ, 493, 222

\bibitem[Millar et al. 1997]{MFW97}  Millar, T. J., Farquhar, P. R. A., \& Willacy, K. 
1997, A\&AS, 121, 139 

\bibitem[Mundy et al. 1985]{MESG85} Mundy, L.G., Evans, N.J. II, Snell, R.L, \& Goldsmith,
P.F. 1985, ApJ, 318, 392

\bibitem[Myers \& Lazarian 1998]{ML98} Myers, P. C. \& Lazarian, A. 1998, in preparation

\bibitem[Myers 1998]{Myers98} Myers, P. C. 1998, ApJ, 496, L109

\bibitem[Myers 1997]{Myers97} Myers, P.C. 1997, in ``Star Formation, Near and Far''
AIP Conference Proceedings 393, eds. S.S. Holt and L.G. Mundy, 41 

\bibitem[Myers, Ladd, \& Fuller 1991]{MLF91} Myers, P.C., Ladd, E.F., \& Fuller, G.A. 1991,
ApJ, 372, L95

\bibitem[Nakano 1998]{Nakano98} Nakano, T.  1998, ApJ, 494, 587.

\bibitem[Neufeld, Lepp, \& Melnick 1995]{NLM95}Neufeld, D., Lepp, S., \& Melnick, G. 1995,
ApJS, 100, 132

\bibitem[Plume et al. 1998]{PBWM98} Plume, R., Bergin, E.A., Williams, J.P., \& Myers,
P.C. 1998, Faraday Discuss., 109, in press

\bibitem[Plume et al. 1997]{Plumeetal97} Plume, R., Jaffe, D.T., Evans, N.J. II, Mart\'in-Pintado, J., \& G\'omez-Gonz\'alez, J. 1997, ApJ, 476, 730

\bibitem[Shu, Adams, \& Lizano 1987]{SAL87} Shu, F. H., Adams, F.C., \& Lizano, S. 1987,
ARA\&A, 25, 23

\bibitem[Snell 1981]{Snell81} Snell, R.L. 1981, ApJS, 45, 121

\bibitem[Strom, Strom, \& Merrill 1993]{SSM93} Strom, K.M., Strom, S.E., \& Merrill, K.M.,
ApJ, 412, 233

\bibitem[Tatematsu et al. 1993a]{Tatematsuetal93a} Tatematsu, K. et al. 1993a, ApJ, 404, 643

\bibitem[Tatematsu et al. 1993b]{Tatematsuetal93b} Tatematsu, K., Umemoto, T., Murata, Y., 
Chen, H., Hirano, N., \& Takaba, H. 1993b, ApJ, 419, 746 

\bibitem[Ungerechts et al. 1997]{FC_OMC197}
Ungerechts, H.A., Bergin, E.A., Goldsmith, P.F., Irvine, W.M., Schloerb, F.P.,
\& Snell, R.L. 1997, \apj, 482, 245

\bibitem[Ward-Thompson et al. 1994]{WSHA94} Ward-Thompson, D., Scott, P.F., Hills, R.E., Andre, P. 1994, MNRAS, 268, 276

\bibitem[Williams et al. 1998]{WBCMP98} Williams, J.P., Bergin, E.A., Caselli, P., 
Myers, P.C., \& Plume, R. 1998, ApJ, in press (Paper I)

\bibitem[Wootten, Loren, \& Snell 1982]{WLS82} Wootten, A., Loren, R.B., \& Snell, R.L. 1982, ApJ, 255, 160

\bibitem[Zavagno et al. 1997]{ZMTSG97} Zavagno, A., Molinari, S., Tommasi, E., Saraceno,
P., \& Griffin, M. 1997, A\&A, 325, 685

\bibitem[Zinnecker, McCaughrean, \& Wilking 1993]{ZMW93} Zinnecker, H., McCaughrean, M.J., \& Wilking, B.A. 1993, in Protostars and Planets III, eds. E. Levy, J.I. Lunine,
\& M.S. Matthews (Tucson: U. of Arizona Press), 429

\end{thebibliography}
\end{document}